\documentclass[aps,pra,twocolumn,amsmath,amssymb,superscriptaddress,floatfix]{revtex4-2}
\usepackage[utf8]{inputenc}
\usepackage[english]{babel}
\usepackage{graphicx}
\usepackage{amsmath}
\usepackage{amsthm} 
\usepackage{amssymb}
\usepackage{amsfonts}
\usepackage{mathtools}
\usepackage{enumerate}
\usepackage{units}
\usepackage{xcolor}
\usepackage{color}
\usepackage{physics}
\usepackage{xcolor}
\usepackage[hidelinks, colorlinks=true,linkcolor=blue]{hyperref}
\usepackage{dsfont}

\usepackage{makecell}

\usepackage[inkscapeformat=png]{svg}

\makeatletter
\newsavebox{\@brx}
\newcommand{\llangle}[1][]{\savebox{\@brx}{\(\m@th{#1\langle}\)}%
  \mathopen{\copy\@brx\kern-0.5\wd\@brx\usebox{\@brx}}}
\newcommand{\rrangle}[1][]{\savebox{\@brx}{\(\m@th{#1\rangle}\)}%
  \mathclose{\copy\@brx\kern-0.5\wd\@brx\usebox{\@brx}}}
\makeatother

\newcommand{\cexpval}[1]{\llangle{#1}\rrangle}
\newcommand{\OD}{D}
\newcommand{\ODi}{D_i}
\newcommand{\ODtot}{D_N}
\newcommand{\Gammatot}{\Gamma_\mathrm{tot}}

\begin{document}

\title{Emergence of unidirectionality and phase separation in optically dense emitter ensembles}

\author{Kasper J. Kusmierek}
\affiliation{Institut f\"ur Theoretische Physik, Leibniz Universit\"at Hannover, Appelstraße 2, 30167 Hannover, Germany}
\author{Max Schemmer}
\affiliation{Istituto Nazionale di Ottica del Consiglio Nazionale delle Ricerche (CNR-INO), 50019 Sesto Fiorentino, Italy}
\affiliation{European Laboratory for Non-Linear Spectroscopy (LENS), Università di Firenze, 50019 Sesto Fiorentino, Italy}
\author{Sahand Mahmoodian}
\affiliation{Institute for Photonics and Optical Sciences (IPOS), School of Physics, The University of Sydney, NSW 2006, Australia}
\affiliation{Centre for Engineered Quantum Systems, School of Physics, The University of Sydney, NSW, 2006, Australia}
\author{Klemens Hammerer}
\affiliation{Institut f\"ur Theoretische Physik, Leibniz Universit\"at Hannover, Appelstraße 2, 30167 Hannover, Germany}
\date{\today}

\begin{abstract}
    The transmission of light through an ensemble of two-level emitters in a one-dimensional geometry is commonly described by one of two emblematic models of quantum electrodynamics (QED): the driven-dissipative Dicke model or the Maxwell-Bloch equations. Both exhibit distinct features of phase transitions and phase separations, depending on system parameters such as optical depth and external drive strength. Here, we explore the crossover between these models via a parent spin model from bidirectional waveguide QED, by varying positional disorder among emitters. Solving mean-field equations and employing a second-order cumulant expansion for the unidirectional model---equivalent to the Maxwell-Bloch equations---we study phase diagrams, the emitter's inversion, and transmission  depending on optical depth, drive strength, and spatial disorder. We find in the thermodynamic limit the emergence of phase separation with a critical value  that depends on the degree of spatial order but is independent of Doppler broadening effects. Even far from the thermodynamic limit, this critical value marks a special point in the emitter's correlation landscape of the unidirectional model and is also observed as a maximum in the magnitude of inelastically transmitted photons. We conclude that a large class of effective one-dimensional systems without tight control of the emitter's spatial ordering can be effectively modeled using a unidirectional waveguide approach.
\end{abstract}

\maketitle

\section{Introduction}
The transmission of light through atomic media is perhaps the quintessential quantum optics experiment. There is a vast array of experimental as well as theoretical works studying this paradigm~\cite{McCall1969PR,  Mahmoodian2018PRL,Veyron_2022, Kusmierek2023SCIPOST,kusmierek_code_2025, agarwal2024, goncalves_driven-dissipative_2024, ferioli_non-equilibrium_2023, FerioliarXiv2024}. Such works have examined atomic correlations that include phase transitions~\cite{ferioli_non-equilibrium_2023,agarwal2024,goncalves_driven-dissipative_2024,ruostekoski_superradiant_2024} and atomic spin squeezing~\cite{Kuzmich1997PRL, Huber2022PRA}, as well as the correlations of the transmitted photons~\cite{Lu1998PRL,Mahmoodian2018PRL, Mahmoodian2020PRX, prasad_correlating_2020, hinney_unraveling_2021,Kusmierek2023SCIPOST,kusmierek_code_2025,cordier_tailoring_2023,schemmer_simple_2024}.

\begin{figure}[t]
    \centering
    \includegraphics[width=\columnwidth]{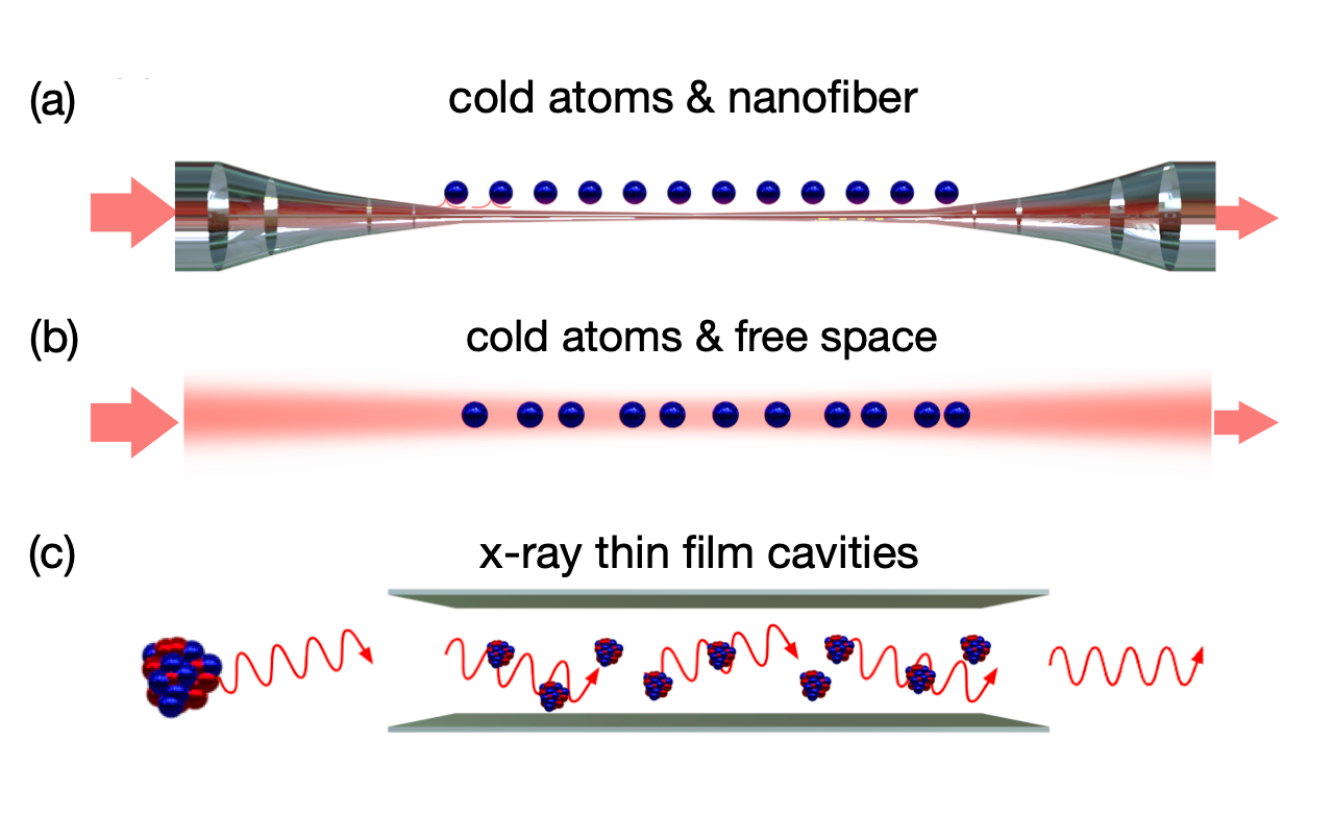}
    \caption{Examples of experimental realizations in which unidirectionality and phase seperation emerge. a) Laser-cooled atoms coupled to a tapered optical nanofibers. b) Laser-cooled atoms in free space, and c) M\"ossbauer nuclei in a thin film cavity.}
\label{fig:schemFirst}
\end{figure}

Three prominent experimental geometries for studying the transmission of light are shown in Fig.~\ref{fig:schemFirst} including cold atoms trapped in the vicinity of an optical nanofiber~\cite{vetsch_optical_2010,corzo_large_2016,sorensen_coherent_2016,nieddu_optical_2016,solano_chapter_2017,prasad_correlating_2020,hinney_unraveling_2021,su_dark_2022,cordier_tailoring_2023,kestler_state-insensitive_2023}, cold atoms trapped in free space~\cite{araujo_superradiance_2016,kashanian_noise_2016,glicenstein_collective_2020,lassegues_transition_2023,ferioli_non-equilibrium_2023,FerioliarXiv2024}, and an ensemble of nuclei coupled to an x-ray waveguide~\cite{salditt_x-ray_2020,lohse_collective_2024}. Although these platforms are experimentally very different, their modeling often coincides, and they can be described with the same minimal theoretical frameworks, such as the Dicke model~\cite{ferioli_non-equilibrium_2023,goncalves_driven-dissipative_2024,agarwal2024}, or a unidirectional (chiral, cascaded) waveguide model~\cite{prasad_correlating_2020,hinney_unraveling_2021,cordier_tailoring_2023,agarwal2024,goncalves_driven-dissipative_2024,ruostekoski_superradiant_2024, lentrodt_excitation_2024}. Interestingly, both models predict a phase transition/separation, with similar features~\cite{gilmore_relation_1978,emary_chaos_2003,goncalves_driven-dissipative_2024,agarwal2024,ruostekoski_superradiant_2024}.
While these minimal models are computationally practical, they may not always transparently reflect experimental conditions. For instance, directional coupling emerges naturally in nanofiber systems~\cite{mitsch_quantum_2014, Lodahl2017Nature} but is typically absent in free-space setups. 

The general many-body dynamics of these systems can be theoretically described using a Lindblad master equation where the coupling between emitters is mediated by the electromagnetic environment. For emitters coupled to waveguides, e.g. the nanofiber and x-ray waveguide, the emitter-emitter coupling is long-range and mediated by the two counterpropagating bound modes of the waveguide. This paradigm of light-matter coupling is described by a many-body bidirectional waveguide master equation \cite{Pichler2015PRA}. 
When neglecting the propagation of the electromagnetic field through the ensemble, i.e. all emitters are coupled to the mode at an identical position,  this reduces to the driven-dissipative Dicke model (DM).
This approximation can also be made in cavity quantum electrodynamics (QED) systems. The DM has a permutationally invariant Hamiltonian whose solutions, in the absence of individual emitter dissipation, decouple into subspaces of fixed total angular momentum enabling simplified numerical and analytic treatments. On the other hand, if the coupling to the left- and right-propagating modes in the bidirectional master equation is not the same, the master equation becomes chiral and in the extreme limit of unidirectional coupling is described by the cascaded master equation \cite{Gardiner1993PRL}. Within the continuum and mean-field limits the cascaded master equation is equivalent to the Maxwell-Bloch equations \cite{McCall1969PR}. These can be solved analytically in the mean-field limit and give rise to the well-known Beer-Lambert law~\cite{bouguer_essai_1729,lambert_photometria_1760,beer_bestimmung_1852}.

Both the Maxwell-Bloch equations and the DM have had success in predicting the transmission properties of quantum optics experiments~\cite{McCall1969PR, ferioli_non-equilibrium_2023,lechner_light-matter_2023,agarwal2024,goncalves_driven-dissipative_2024,ruostekoski_superradiant_2024}. However, the regime of validity of these models and their relation to each other has not always been obvious. For example, disordered atomic ensembles in free space such as shown in Fig.~\ref{fig:schemFirst}(b), have a combination of atom-atom interactions that are short range $\propto 1/r^3$ and long range $\propto 1/r$. Nevertheless, these systems have been modeled extensively using the DM or Maxwell-Bloch equations \cite{McCall1969PR, goncalves_driven-dissipative_2024} which tend to give good agreement with free-space models \cite{agarwal2024}. However, the DM is not generic and requires equal magnitude and vanishing propagation phase for all emitter-light interactions. Reaching precisely these conditions is difficult for many experimental systems because of imperfections and disorder.

Here, we present a comprehensive analysis of the transmission properties of a coherently driven ensemble of two-level emitters coupled to a one-dimensional continuum of forward and backward propagating modes, examining both the observables of the emitters and the electromagnetic field.
We find that by varying disorder, within this bidirectional waveguide model, a transition occurs from dissipative Dicke dynamics to unidirectional dynamics. We show this analytically via an ensemble average and numerically for single realizations using mean-field theory. 

Equipped with our bidirectional waveguide model, we examine the saturation dynamics of the ensemble with varying optical depth and input laser power. Previous work studying models with unidirectional coupling \cite{Mahmoodian2018PRL} showed that, at large optical depths and weak saturation power, rather than observing exponential decay of transmitted power versus optical depth associated with the Beer-Lambert law, the system exhibits a power-law decay scaling as $1/D^{3/2}$, where $D$ is the optical depth. This work hinted at the presence of critical dynamics in the driven ensemble. More recently, the presence of a non-equilibrium phase separation was observed in a driven free-space atomic ensemble \cite{ferioli_non-equilibrium_2023}. In that work, the ensemble was driven by a resonant laser with power larger than the saturation power of a single atom. The atoms at the front of the ensemble are saturated and in a mixed state, while the back of the array has atoms in the ground state.  Surprisingly, the transition between these two regimes is sharp, separating the ensemble into two distinct phases.  Although this was originally modeled using the dissipative DM, very recently, it has been shown that the phase separation is also a feature of the Maxwell-Bloch equation \cite{goncalves_driven-dissipative_2024}, as well as free-space models \cite{agarwal2024, ruostekoski_superradiant_2024}. 

Here, we specifically investigate the role of a phase separation in the saturation dynamics of a driven ensemble in the presence of disorder in a system with bidirectional coupling.  At two extremities of no disorder and large disorder, corresponding to DM dynamics and unidirectional dynamics respectively, the phase separation occurs at different positions in the array. In addition to this, by slowly varying the disorder parameter we observe what appears to be a transition between the ordered phase separation and disordered phase separation. This transition is accompanied by large fluctuations in the emitters' inversion and in the transmitted fields. Finally, we push our theoretical description of the unidirectional model beyond mean-field theory and compute the emitter's two-body correlation functions and the inelastic component of the transmitted field. For the emitter's correlators, the phase coexistence separates regions of strong correlations with regions of no correlations. In case of the transmitted field, the phase separation coincides with a maximum in the inelastically transmitted photons and, with that, in the integrated in-phase and out-of-phase quadrature fluctuations.

This paper is organized as follows: In Sec.~\ref{sec:1DQED} we introduce the description of an ensemble coupled to a one-dimensional continuum by the bidirectional master equation. In Sec.~\ref{sec:Crossover}, we show how the bidirectional master equation leads to the driven-dissipative Dicke model and the unidirectional waveguide model in limits of no disorder and large disorder. Then in Sec.~\ref{sec:Numerics} we provide numerical results for the inversion, transmission and reflection in the ensemble for different values of disorder as well as the two-body correlation functions for the unidirectional system. We also consider the influences of Doppler broadening on the phase separation in the unidirectional model. Finally in Sec.~\ref{sec:summaryAndOutlook}, we provide a summary and outlook and compare the predictions of our model with recent works.

\section{Master equation for 1D QED}\label{sec:1DQED}
In its most simple form the transmission of light through an ensemble of two-level scatterers is realized and studied in a one-dimensional geometry.
In this section, we introduce the description of such an optically dense ensemble coupled to the EM field free to propagate in either direction along this dimension. The master equation and the physics implied by it will depend sensitively on the positions of the emitters as will be explored in detail in the next Secs.~\ref{sec:Crossover} and \ref{sec:Numerics}.

\subsection{Master Equation}

The dynamics of $N$ emitters coupled bidirectionally with a rate $\Gamma_{\rm 1D}$ to a one-dimensional waveguide is described under the Markov and rotating wave approximation by the master equation (MEQ)~\cite{Pichler2015PRA} 
\begin{equation}\label{eq:Bi-MEQ}
\begin{split}
&\frac{d \hat{\rho}}{d t} = -i \left[\hat{H}_{\rm eff}, \hat{\rho} \right] + \sum_{i=1}^N  \frac{\gamma}{2} \left[2 \hat{\sigma}_i^- \hat{\rho} \hat{\sigma}_i^+ - \hat{\sigma}^+_i \hat{\sigma}^-_i \hat{\rho} - \hat{\rho} \hat{\sigma}^+_i \hat{\sigma}^-_i \right]\\
+& \sum_{i,j=1}^N \frac{\Gamma_{\rm 1D}}{2} \cos{\left( k_0 |z_i - z_j| \right)} \left[2 \hat{\sigma}_i^- \hat{\rho} \hat{\sigma}_j^+ - \hat{\sigma}^+_i \hat{\sigma}^-_j \hat{\rho} - \hat{\rho} \hat{\sigma}^+_i \hat{\sigma}^-_j \right],
\end{split}
\end{equation}
with 
\begin{equation}
\begin{split}
\hat{H}_{\rm eff} &=  -\sum_{i=1}^N \Delta \hat{\sigma}_i^+ \hat{\sigma}_i^- + \frac{\Omega_i}{2} \hat{\sigma}^-_i + \frac{\Omega_i^*}{2} \hat{\sigma}^+_i \\
&+ \sum_{i,j=1}^N \frac{\Gamma_{\rm 1D}}{2} \sin{\left( k_0 |z_i - z_j| \right)} \hat{\sigma}^+_i \hat{\sigma}^-_j,
\end{split}
\end{equation}
written in a frame rotating at the frequency of the laser $\omega$ such that $\Delta=\omega-\omega_0$ is the detuning between the laser and the emitter's resonance.  
Here, $\hat{\sigma}_{i}^-$ ($\hat{\sigma}_{i}^+$) is the two-level lowering (raising) operator for the $i$--th emitter and $\Omega_i  = -  2\langle \hat{\mathbf E}_{\rm in}^{-}(\mathbf r_i)\rangle \cdot \mathbf d$ is the Rabi frequency due to the coherent driving field of the $i$--th emitter, where $\mathbf d$ is the transition dipole moment of the two-level emitters. We assume the decay rates into the forward and backward propagating modes are both $\Gamma_{\rm 1D}/2$. Thus, we do not assume an intrinsic chirality in the light-matter interaction. The coupling to forwards and backwards modes is taken to be equal and  each emitter is also coupled to an external loss channel with a rate $\gamma$, such that the total decay rate is $\Gammatot=\Gamma_\mathrm{1D}+\gamma$. 

The assumptions under which MEQ~\eqref{eq:Bi-MEQ} can be derived from the general 3D light-matter problem is provided in Appendix~\ref{sec:app_MEQ}. 

\begin{figure}[t]
        \centering 
        \includegraphics[width=\linewidth]{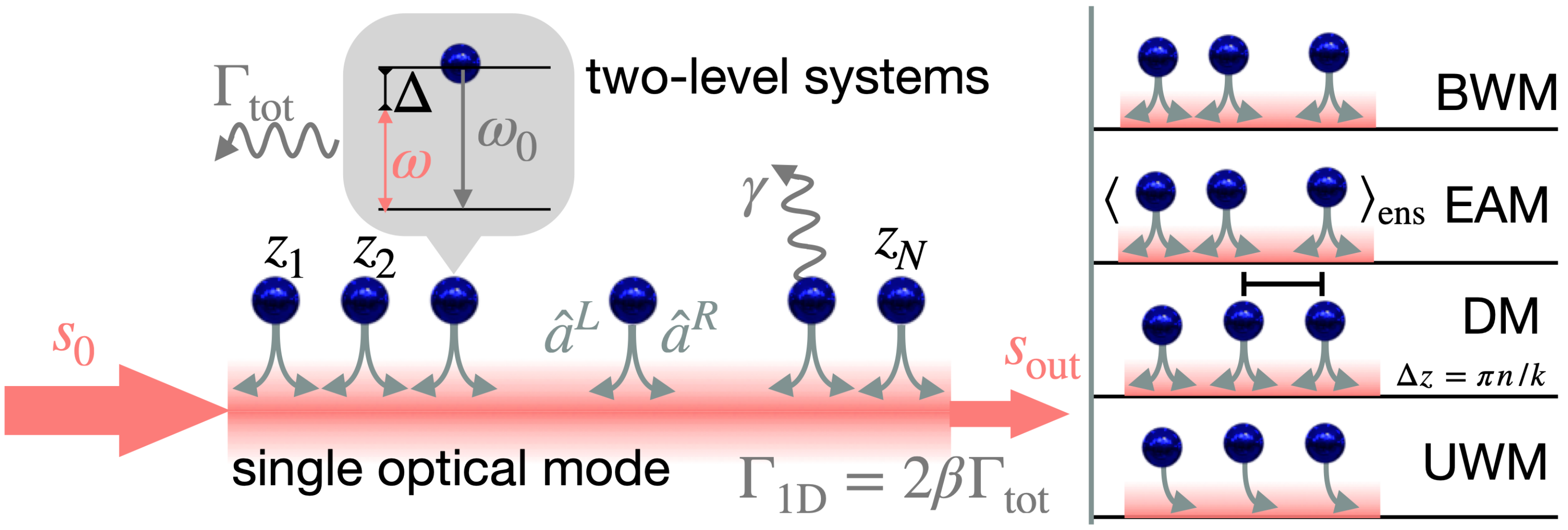}
    \caption{Left: Waveguide model describing the experimental systems in Fig.~\ref{fig:schemFirst} (a) - (c). Right: Sketches of the four different theoretical model used in this article to model the situation on the left. Bidirectional waveguide model (BWM), ensemble averaged waveguide model (EAM), where $\langle  \ldots \rangle_\mathrm{ens}$ denotes ensemble averaging, the driven-dissipative Dicke model (DM), and the unidirectional waveguide model (UWM), details see Table~\ref{tab:models}.}
        \label{fig_model}
        \end{figure}

\subsection{Input-output relations and Rabi frequency}

In order to use this master equation to calculate expectation values of the transmitted and reflected field one also needs input-output relations. From the expression for the electric field operator (\ref{eq:electricFieldOperator}) and Green's tensor (\ref{eq:greensTensor}) we have
\begin{align}
\label{eq:inputOutputRelations1}
\hat{a}^R_{\rm out}(t) = \hat{a}^R_{\rm in}(t) - i \sqrt{\frac{\Gamma_{\rm 1D}}{2 v_g}} \sum_{i=1}^N  \hat{\sigma}_i^-(t) e^{i k_0 (z_N - z_i)},\\
\label{eq:inputOutputRelations2}
\hat{a}^L_{\rm out}(t) = \hat{a}^L_{\rm in}(t) - i \sqrt{\frac{\Gamma_{\rm 1D}}{2 v_g}} \sum_{i=1}^N \hat{\sigma}_i^-(t) e^{i k_0 (z_i - z_1)}.
\end{align}
Here $\hat{a}_{\rm out}^R$ ($\hat{a}_{\rm out}^L$) gives the annihilation operator for the right-propagating (left-propagating) input waveguide field just to the right (left) of the $N$-th (first) emitter in the ensemble, and $\hat{a}_{\rm in}^R$ ($\hat{a}_{\rm in}^L$) gives the right-propagating (left-propagating) input-field annihilation operator. A detailed derivation of the input-output relations is provided in Appendix \ref{sec:app_inputOutput}. By combining the forwards and backwards propagating fields we obtain an expression for the photon operator at any position $z$ along the propagation direction
\begin{equation}
\label{eq:fieldOperatorInWaveguide}
\begin{split}
\hat{a}(z, t) =&   \hat{a}^R(z - v_g t, 0) + \hat{a}^L(z + v_g t, 0)\\
&- i \sqrt{\frac{\Gamma_{\rm 1D}}{2 v_g}} \sum_{i=1}^N \hat{\sigma}^-_i(t) e ^{i k_0 |z - z_i|} .
\end{split}
\end{equation}
Here $\hat{a}^{R/L}(z \mp v_g t, 0)$ are the right/left photon annihilation operators acting on the photon state of the waveguide at at time $t=0$. This expression is used to obtain the expectation values for the fields between emitters along the ensembles. 

Finally, we need to relate the Rabi frequency $\Omega_i$ in the master equation to the input field operators.  We assume that the ensemble is driven by a coherent monochromatic right-propagating field that is in the waveguide while all other fields are initially in the vacuum state. We obtain for the Rabi frequency (see Appendix \ref{sec:app_inputOutput}),
\begin{align}
\label{eq:RabiFreq}
    \Omega_i = 2 \sqrt{\frac{\Gamma_{\rm 1D} v_g}{2}} \expval*{ \hat{a}^{R \, \dagger}_{\rm in}} e^{-i k_0 z_i},
\end{align}
where $\abs*{\langle \hat{a}^{R \, \dagger}_{\rm in}\rangle}^2$ denotes the input photon flux from the left, cf. Fig. \ref{fig_model}. The effective spin model given by Eqs.~\eqref{eq:Bi-MEQ} and \eqref{eq:fieldOperatorInWaveguide} provides the basis for the next two sections where we will study certain limiting cases resulting from specific choices of the emitter's positions $z_i$. In particular, we demonstrate the emergence of the driven-dissipative Dicke model and the unidirectional waveguide model and investigate the crossover between them.
\section{Limits of dissipative Dicke model and unidirectional waveguide model}\label{sec:Crossover}
%
The physics of a 1D bidirectional system, as described by the master equation~\eqref{eq:Bi-MEQ}, is very sensitive to the positions and ordering of the emitters. 
A special case is the configuration of a so-called Bragg-reflector~\cite{Chang2012NJP, Chen2012}, where emitters are positioned regularly with a spacing equal to multiples of $\lambda/2$. In this limit, the spin model~\eqref{eq:Bi-MEQ} amounts to the famous driven-dissipative Dicke model~\cite{Kirton_2018}, as was shown in~\cite{Pichler2015PRA} and will be discussed below. Taking this specific configuration as a reference point, we then add disorder to the emitter's positions and show that, upon ensemble averaging, Eq.~\eqref{eq:Bi-MEQ} maps to a unidirectional master equation~\cite{Stannigel2012NJP,Pichler2015PRA,Kusmierek2023SCIPOST} in the limit of large disorder.
These limiting cases and their models of complete positional order and large disorder are well discussed in the literature. However, since we will treat the intermediate regime of finite disorder by a smooth interpolation between these models and their vastly different physics and phases, we discuss in this section some of their properties that will be important in the following section.
In order to account for disorder it will be useful to reformulate the spin model of Eq.~\eqref{eq:Bi-MEQ} such that it only depends on relative rather than absolute positions (as is the case in Eq.~\eqref{eq:RabiFreq}). A reformulation of this form is possible in a suitable gauge, referred to as spiral gauge in~\cite{agarwal2024}, as we discuss in the following section.

\subsection{Bidirectional waveguide model in spiral gauge and ensemble average}

The position dependence is encoded in the phases in the general master equation \eqref{eq:Bi-MEQ}. To simplify calculations we apply a local gauge transformation $\hat{\sigma}_i^-\rightarrow\hat{\sigma}_i^-e^{ik_0z_i}$ to \eqref{eq:Bi-MEQ} to absorb the driving phase. For the field operators, this gauge means $\hat{a}(z,t)\rightarrow\hat{a}(z,t)e^{ik_0z}$. Further, we set from here on $\Delta=0$ and $v_g=1$. The explicit form of the master equation in spiral gauge, whose physics is of course equivalent to the one of Eq.~\eqref{eq:Bi-MEQ}, is given in Appendix~\ref{app:ensemble_average} in Eq.~\eqref{eq:MEQ_spiral}. In the following we refer to the resulting model as bidirectional waveguide model (BWM).

The fact that Eq.~\eqref{eq:MEQ_spiral} does not depend on absolute positions can be exploited to average the master equation. The position of the emitters are sampled by treating the relative distance $z_{j+1}-z_j$ of neighboring emitters as independent random variables described by a Gaussian distribution. For the Gaussian distribution, we choose a mean $ \frac{\lambda}{2}$, corresponding to a Bragg-reflector configuration, and a standard deviation of $\frac{\lambda}{2}\eta$, where $\eta\in[0,\infty)$ parametrizes the degree of disorder.
To be able to compare different regimes of the system with respect to $\eta$, we average the BWM over the positions of the emitters 
We refer to Appendix \ref{app:ensemble_average} for a detailed calculation. The corresponding ensemble averaged model (EAM) reads 
\begin{align}\label{eq:Ens-Ave_MEQ}
    \frac{d\hat{\rho}}{dt}=&-i\qty[\hat{H}_\mathrm{sys}+\hat{H}_L^\mathrm{av}+\hat{H}_R,\hat{\rho}]\nonumber\\
    &+ \sum_{ij}\Gamma_{ij}^\mathrm{av}\qty(\hat{\sigma}^-_i\hat{\rho}\hat{\sigma}^+_j - \frac{1}{2}\hat{\sigma}^+_i\hat{\sigma}^-_j\hat{\rho} - \frac{1}{2}\hat{\rho}\hat{\sigma}^+_i\hat{\sigma}^-_j)
\end{align}
with
\begin{align}\label{eq:H_sys}
    \hat{H}_\mathrm{sys} & = \sum_j\frac{\Omega}{2}\qty( \hat{\sigma}^-_j + \hat{\sigma}_j^{+}) \\
    \label{eq:H_R}
    \hat{H}_R & =-\frac{i \Gamma_\mathrm{1D}}{4} \sum_{j>i}\left(\hat{\sigma}_j^{+} \hat{\sigma}_i^--\mathrm{ h.c. }\right) \\
    \hat{H}_L^\mathrm{av} & =-\frac{i \Gamma_\mathrm{1D}}{4} \sum_{j<i}e^{-2(\eta\pi)^2\abs{i-j}}\left( \hat{\sigma}_j^{+}\hat{\sigma}_i^--\mathrm{ h.c. }\right) 
\end{align}
and
\begin{align}
    \Gamma_{ij}^\mathrm{av}= \gamma\delta_{ij} + \frac{\Gamma_\mathrm{1D}}{2}\qty(1+e^{-2(\eta\pi)^2\abs{i-j}}).
\end{align}
Since the phases only enter in describing the left-propagating mode, the corresponding field at position $z_i$ takes the form 
\begin{align}\label{eq:Ens-Ave_in-out}
\hat{a}^\mathrm{EAM}(z_i,t)=&\hat{a}_{\mathrm{in}}(z_i,t)-i \sqrt{\frac{\Gamma_{1 \mathrm{D}}}{2}} \sum_{j: z_j<z_i} \hat{\sigma}_j^{-}(t)\nonumber \\ &-i  \sqrt{\frac{\Gamma_{1 \mathrm{D}}}{2}}\sum_{j: z_j>z_i} \hat{\sigma}_j^{-}(t) e^{-2(\eta\pi)^2\abs{i-j}}
\end{align} 
where we can define $\hat{a}_\mathrm{in}(z_i,t)=\hat{a}^R(z_i-t,0)+\hat{a}^L(z_i+t,0)$, reflecting that we always assume a drive from the left and vacuum input from the right, cf. Fig.~\ref{fig_model}a \footnote{We note that Eq. \eqref{eq:Ens-Ave_in-out} can only be used to compute the coherent part of the field. For the total photon flux, which incorporates the incoherent part as well, one needs to compute this quantity first for a single realization and then take the average over the ensemble.}. 

The computational complexity of the inhomogeneous spin model in Eq.~\eqref{eq:Ens-Ave_MEQ} 
rises, in general, exponentially. As an approximate treatment in the high particle limit we consider the mean-field equations, which read for the EAM,
\begin{subequations}
\label{eq:MF}
\begin{align}\label{eq:MF_sigmin}
\frac{d\expval*{\hat{\sigma}_{i}^{-}}}{dt}
&=i\alpha_i\expval*{\hat{\sigma}_i^z}-\frac{\Gammatot}{2}\expval*{\hat{\sigma}_i^-}\\
 \frac{d\expval*{\hat{\sigma}_i^z }}{dt} &=2i\alpha_i^*\expval*{\hat{\sigma}_i^-}-2i\alpha_i\expval*{\hat{\sigma}_i^+} -\Gammatot\left(1+\expval*{\hat{\sigma}_i^z}\right).\label{eq:MF_sigz}
 \end{align}
\end{subequations}
The effective Rabi frequency of emitter $i$ is
\begin{align}\label{eq:effectiv_drive_EnsAve}
\alpha_i^\mathrm{EAM} =& \frac{\Omega}{2} -i\frac{\Gamma_\mathrm{1D}}{2}\sum_{j=1}^{i-1}\expval*{\hat{\sigma}_j^-}\nonumber\\
&-i\frac{\Gamma_\mathrm{1D}}{2}\sum_{j=i+1}^{N}e^{-2(\eta\pi)^2\abs{i-j}}\expval*{\hat{\sigma}_j^-}.
\end{align}
These equations follow from Eq.~\eqref{eq:Ens-Ave_MEQ} in the mean-field approximation of factorizing two-particle correlations $\expval*{\hat{\sigma}_i^\alpha\hat{\sigma}_j^\beta}=\expval{\hat{\sigma}_i^\alpha}\expval*{\hat{\sigma}_j^\beta}$. Here, as in every other model in this article, the relation between the spatially varying field operator and the effective Rabi frequency is $\sqrt{\frac{\Gamma_\mathrm{1D}}{2}}\expval*{\hat{a}(z_i,t)}=\alpha_i$.

We will also consider a mean-field Ansatz for the BWM in Eq.~\eqref{eq:MEQ_spiral} for single realizations of emitter's positions, drawn according to the Gaussian statistics assumed for the EAM. The mean-field equations for the BWM are identical to Eqs.~\eqref{eq:MF}, but with different effective Rabi frequencies $\alpha_i$, as given in~\eqref{eq:effectiv_drive_Pos}. For convenience and easier reference, we collect the main equations (Master Equation, input-output relations, and mean-field equations) of the EAM and BWM in Table ~\ref{tab:models}. The solution to the mean-field equations will be discussed in Sec.~\ref{sec:Numerics}. 
Before that, we consider the limiting cases of $\eta\rightarrow 0$ and $\eta\rightarrow\infty$ of the EAM which yield, respectively, the driven-dissipative Dicke and the unidirectional waveguide model, and review some of their properties that will be important to the discussion ahead.

\subsection{Driven-dissipative Dicke model}

        \begin{table}[t]
    \centering
        \begin{tabular}{|cc|c|c|c|}
            \hline
             & & MEQ & \makecell{in-out\\relations} & \makecell{MF\\equations} \\ \hline\hline
             BWM & \makecell{ Bidirectional \\ waveguide model} & 
             \eqref{eq:MEQ_spiral} & 
             \eqref{eq:spiral_in-out} & 
             \eqref{eq:MF} \eqref{eq:effectiv_drive_Pos} \\[10pt]
             
             EAM & \makecell{Ensemble averaged \\ waveguide model, $\eta$}   & 
             \eqref{eq:Ens-Ave_MEQ} & 
             \eqref{eq:Ens-Ave_in-out} & 
             \eqref{eq:MF} \eqref{eq:effectiv_drive_EnsAve} \\[10pt]
             
             DM & \makecell{Driven-dissipative \\ Dicke model, $\eta\rightarrow 0$}   & 
             \eqref{eq:Dicke_model} & 
             \eqref{eq:Dicke_in-out} & 
             \eqref{eq:MF} \eqref{eq:effectiv_drive_Dicke} \\[6pt]
             
             UWM & \makecell{Unidirectional \\waveguide model,\\ $\eta\rightarrow\infty$} & 
             \eqref{eq:Chiral_MEQ} & 
             \eqref{eq:Chiral_in-out} & 
             \eqref{eq:MF} \eqref{eq:effectiv_drive_Chiral} \\ \hline
        \end{tabular}
        \caption{
        Overview of models that are used to describe the situation in Fig.~\ref{fig_model}. BWM: A single realization of a bidirectional waveguide system. EAM: A position independent average over an ensemble of BWM systems. DM: EAM taken to the limit of $\eta\rightarrow 0$. UWM: EAM taken to the limit of $\eta\rightarrow\infty$. Abbreviations on top: Master equation (MEQ), input-output relation (in/out) and mean-field equation (MF).}
\label{tab:models}
\end{table}

For $\eta=0$ the emitters are in complete order with a fixed distance obeying the Bragg condition. In this case, the light field acquires no net phase traveling from one emitter to the other. Since the information of the individual emitter's positions drop out, the system is permutationally symmetric, and the BWM in Eq.~\eqref{eq:MEQ_spiral} reduces to  
\begin{align}\label{eq:Dicke_model}
    \frac{d\hat{\rho}}{dt}&=-i\frac{\Omega}{2}\sum_j\left[\left( \hat{\sigma}^-_j+ \hat{\sigma}_j^{+}\right), \hat{\rho}\right]\nonumber\\
    &\quad+\Gamma_\mathrm{1D} \mathcal{D}\left[\sum_j\hat{\sigma}_j^-\right] \hat{\rho}+\gamma \sum_j\mathcal{D}[\hat{\sigma}_j^-]\hat{\rho}
\end{align}
    with $\mathcal{D}\left[x\right]\hat{\rho}=x\hat{\rho}x^\dagger-\frac{1}{2}(x^\dagger x\hat{\rho}+\hat{\rho}x^\dagger x)$,
and the field along the axis is given by
\begin{align}\label{eq:Dicke_in-out}
    \hat{a}^\mathrm{DM}(z,t)=\hat{a}_{\mathrm{in}}(z,t)-i \sqrt{\frac{\Gamma_ \mathrm{1D}}{2}} \sum_{j: z_j\neq z} \hat{\sigma}_j^{-}(t) .
\end{align}
We refer to the model described by \eqref{eq:Dicke_model} and \eqref{eq:Dicke_in-out} as the driven-dissipative Dicke model. Up to the Lindblad terms accounting for individual decay, this model is also referred to as Cooperative-Resonance-Fluorescence model in \cite{agarwal2024}. The driven-dissipative Dicke model is usually considered and well established in the context of cavity QED where an ensemble of emitters is coupled to one electromagnetic field mode~\cite{Kirton_2018}. The DM also emerges if a 1D ensemble distributed along a waveguide with a fixed emitter distance satisfies the Bragg condition, as was observed already in~\cite{Pichler2015PRA}. This has drastic consequences regarding the computational complexity, which grows as $\mathcal{O}(N^3)$ due to the permutational symmetry, rather than exponentially as for non-symmetric systems (such as the BWM or EAM)~\cite{Chase_2008, Baragiola_2010,Shammah_2018}, cf. also earlier work concerning the DM~\cite{Sarkar_1987_I, Sarkar_1987_II}. Additionally, mean-field descriptions of these types of systems become exact in the high-particle limit. Taking advantage of this, we can concentrate our analysis on the equations of motion in mean-field approximation for emitter's variables in Eq.~\eqref{eq:MF} with the effective Rabi frequency
\begin{align}\label{eq:effectiv_drive_Dicke}
\alpha^\mathrm{DM} = \frac{\Omega}{2} -i\frac{\Gamma_\mathrm{1D}}{2}(N-1)\expval*{\hat{\sigma}^-}.
\end{align}
Here, the particle index $n$ of both the field $\alpha^\mathrm{DM}$ and the emitter's dipole $\expval*{\hat{\sigma}^-}$ has been dropped, exploiting the permutational symmetry. The complexity of these equations lies solely in the non-linearity of the equations of motion, which implies the possibility of multistability.

Since we are interested in the steady-state solution, we insert the effective Rabi frequency \eqref{eq:effectiv_drive_Dicke} into Eq.~\eqref{eq:MF}, and solve \eqref{eq:MF_sigmin} for $\langle\hat{\sigma}^-_i\rangle$. Using the solution in \eqref{eq:MF_sigz} we arrive at a cubic equation \cite{Carmichael_1980,leppenen2024} for the steady state inversion $\langle\hat{\sigma}^z\rangle$, 
\begin{align}\label{eq:Dicke_cubic}
    0=&\expval{\hat{\sigma}^z}^3\frac{\ODtot^2}{4}+\expval{\hat{\sigma}^z}^2\left(\frac{\ODtot^2}{4}-\ODtot\right)\nonumber\\
    &+ \expval{\hat{\sigma}^z}\left(s_0-\ODtot+1\right)+1.
\end{align}

We take the opportunity to introduce here a number of dimensionless parameters which will be important for the discussions to come: The optical depth of the entire ensemble of $N$ emitters is $\ODtot=4\beta N$ where $\beta=\Gamma_\mathrm{1D}/(2\Gammatot)$ denotes the fraction of the emitter's emission into the waveguide to the right or to the left, cf. Fig.~\ref{fig:phasesep}. We also used the saturation parameter $s_0=2\Omega
^2/\Gammatot^2$ describing the intensity of the driving field. Equation \eqref{eq:Dicke_cubic} can have up to three distinct real solutions in dependence on the chosen parameters. The system becomes bistable, having one unstable and two stable solutions, for $\ODtot>16$ and an input intensity with saturation parameter  $s_0\in[s^\mathrm{crit}_-,s^\mathrm{crit}_+]$, where
\begin{align}
    s^\mathrm{crit}_{\mp}&=\frac{1}{32}\left(-32\mp\sqrt{\ODtot(\ODtot-16)^3}+\ODtot(40+\ODtot)\right)\\
    &\simeq \begin{cases}
        2\ODtot-4-\frac{8}{\ODtot}\\
        \frac{\ODtot^2}{16}+\frac{\ODtot}{2}+2+\frac{8}{\ODtot}.
    \end{cases}
\end{align}
In the last line, we considered the thermodynamic limit of large system size, that is, the limit of large optical depth $\ODtot$. Thus, in this limit, the bistable region is starting at input saturations $\tilde{s}\geq 2$ where we introduce, also for later use, the scaled saturation parameter $\tilde{s}=s_0/\ODtot$. This type of bistability and hysteresis phenomenon has been regularly observed in experimental Cavity-QED systems, and similar non-linear systems~\cite{BaumannPHDThesis,armen_spontaneous_2009, beaulieu_criticality-enhanced_2025, beaulieu_observation_2025, sett_emergent_2024, fitzpatrick_observation_2017}. For the thermodynamic behavior of this system, the parameter $\tilde{s}$ plays the role of an intensive control parameter. It is simple to check that $\tilde{s}=2\abs*{\langle \hat{a}^{R \, \dagger}_{\rm in}\rangle}^2/N\Gammatot$, which corresponds to the ratio between the input photon flux and the photon flux radiated by a completely saturated ensemble of $N$ emitters into $4\pi$, as pointed out in~\cite{goncalves_driven-dissipative_2024}.
\subsection{Unidirectional waveguide model}\label{sec:chiral_limit}
The second limiting case is achieved by letting $\eta\rightarrow\infty$, i.e. the emitters are randomly distributed along the chain with no fixed relation between the positions of neighboring emitters. The phases acquired by light traveling from one emitter to another are random. Along the whole chain, the effect of these random phases cancels. This leaves the right-propagating, forward scattered field unchanged, but prohibits a coherent build-up of a left-propagating mode due to phase mismatch. The left-propagating mode then plays the role of an additional individual decay channel. These statements apply to both the BWM and the EAM. 

For the BWM with fixed and sufficiently random positions of emitters this will become evident in the numerical solutions of the mean-field equations studied in Sec.~\ref{sec:Numerics}. For the EAM, the limit $\eta\rightarrow\infty$ can be easily taken in Eq.~\eqref{eq:Ens-Ave_MEQ}, and yields the unidirectional waveguide model (UWM)
\begin{align}\label{eq:Chiral_MEQ}
    \frac{d\hat{\rho}}{dt}=&-i\left[\hat{H}_{\mathrm{sys}}+\hat{H}_R, \hat{\rho}\right]\\
    &+\frac{\Gamma_\mathrm{1D}}{2} \mathcal{D}\bigg[\sum_j\hat{\sigma}_j^-\bigg] \hat{\rho}\nonumber+\left(\frac{\Gamma_\mathrm{1D}}{2}+\gamma\right)
    \sum_j\mathcal{D}[\hat{\sigma}_j^-]\hat{\rho}
    \end{align}
    with $\hat{H}_\mathrm{sys}$ and $\hat{H}_R$ defined as in Eq.~\eqref{eq:H_sys} and \eqref{eq:H_R}
and the field at position $z$ in the chain
    \begin{align}\label{eq:Chiral_in-out}
    \hat{a}^\mathrm{UWM}(z,t)=\hat{a}_{\mathrm{in}}(t)-i \sqrt{\frac{\Gamma_{1 \mathrm{D}}}{2}} \sum_{j: z_j<z} \hat{\sigma}_j^{-}(t)
    \end{align}
describing only a right-propagating field \cite{Pichler2015PRA,Stannigel2012NJP,Kusmierek2023SCIPOST,kusmierek_code_2025}. 

The computational complexity of the UWM rises exponentially as in the BWM, since the system is not permutationally symmetric. Nevertheless, the unidirectional nature of the interaction implies that mean-field approximations of the master equation~\eqref{eq:Chiral_MEQ} yield coupled sets of \textit{linear} differential equations~\cite{Kusmierek2023SCIPOST,kusmierek_code_2025}. This is in stark contrast to general inhomogeneous models where mean-field approximations generically give rise to highly non-linear equations, suffering from numerical instabilities and stiffness. 
The emerging linearity for the case of the UWM can be seen explicitly by plugging the effective Rabi frequency implied by Eq.~\eqref{eq:Chiral_in-out}
\begin{align}\label{eq:effectiv_drive_Chiral}
	\alpha_i^\mathrm{UWM} = \frac{\Omega}{2} -i\frac{\Gamma_\mathrm{1D}}{2}\sum_{j=1}^{i-1}\expval*{\hat{\sigma}_j^-},
\end{align} 
into the mean-field equations \eqref{eq:MF}. The expectations of the $i$-th emitter couple to themselves and to moments of emitters to the left. But since the moments of the first $i-1$ emitters are independent of the $i$-th one, we can solve for them first and insert the solutions into the equations for the $i$-th emitter. This structure is repeating for higher truncation orders as well, such that the equations for moments of all orders are effectively linear (cf. Appendix in \cite{Kusmierek2023SCIPOST}). 
This effective linearity holds for lowest-order mean-field approximations (where two-particle correlators are factorized), but more generally also for any higher order cumulant expansions (CE), where $k$-particle correlators are approximated in terms of lower order ones. Applying the CE method the complexity grows with the particle number $N$ and truncation order $k$ as 
\begin{align}
    \mathcal{C}=\sum_{i=1}^k 3^i\binom{N}{i},
\end{align}
since this is the number of moments which need to be solved for. Although the system sizes that can be efficiently simulated are still limited in $N$, $\mathcal{C}$ does not grow exponentially. This allows us to employ the CE method in the unidirectional waveguide model and go to higher particle numbers at better approximations than in the BWM and EAM, where the equations can become stiff and numerical approaches suffer from instabilities.

\begin{figure}[t]
    \centering
    \includegraphics[width=\columnwidth]{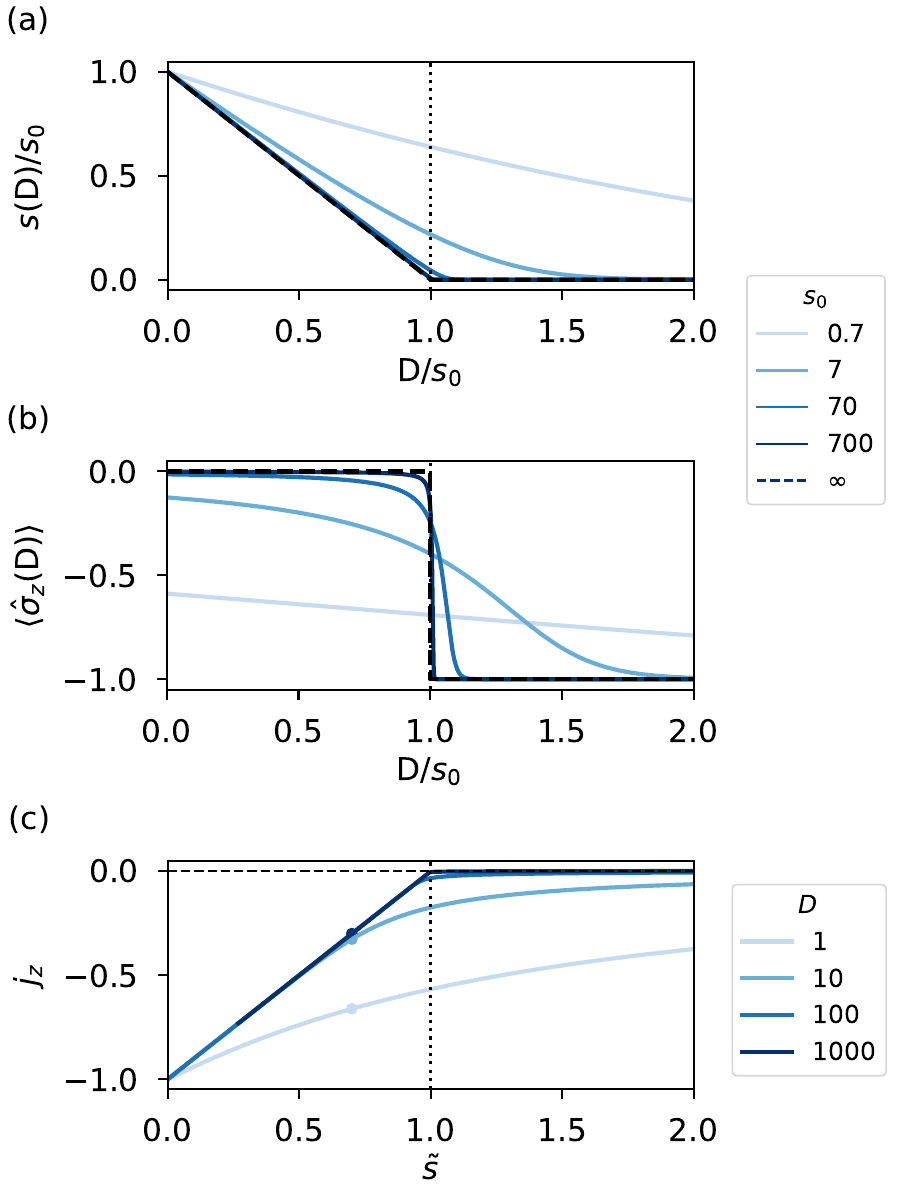}
    \caption{First-order observables of a unidirectional waveguide system exhibiting a phase separation. (a) Elastically scattered light $s(\OD)$ along the chain scaled to the input saturation $s_0$ versus the inverse control parameter $\OD/s_0$. (b) The emitter's inversion  $\expval{\hat{\sigma}_z(\OD)}$ along the chain versus the inverse control parameter $\OD/s_0$. (c) The mean polarization $j_z$ versus the control parameter $\tilde{s}$ for a fixed optical depth. The points indicate the input parameters for which in (a) and (b) the evolution along the chain is shown.}
    \label{fig:phasesep}
\end{figure}

As in the DM, we attempt to find an analytical solution for the mean-field equations \eqref{eq:MF} in steady state. For the UWM, this is done most easily by considering the saturation parameter of the $i$--th emitter $s_i=8\alpha_i^2/\Gammatot^2$. We show in Appendix~\ref{app:Continuos-limit} that Eq.~\eqref{eq:effectiv_drive_Chiral} implies a recursion relation for $s_i$, which can be solved analytically~\cite{Kusmierek2023SCIPOST}. In the UWM, the right-propagating field experiences a growing optical depth $0\leq \ODi\leq 4\beta N$ with $D_i=4\beta i$. In a continuum limit, it is therefore convenient to replace the discrete particle index by a continuous parameter, $i\rightarrow D/4\beta$. The recursion relation for the saturation parameter then takes the simple form
\begin{align}\label{eq:saturation_MF}
    \frac{ds(\OD)}{d\OD}=-\frac{s(\OD)}{1+s(\OD)}.
\end{align}
Note the different interpretation of the optical depth in the DM and UWM. Since the former is permutationally symmetric, only the total emitter number $N$ and net optical depth $\ODtot=4\beta N$ are of importance. In contrast, in the UWM $\ODi$ and $\OD$ can be used effectively as a coordinate along the axis of propagation. 

It is instructive to rewrite Eq.~\eqref{eq:saturation_MF} in real space assuming a constant number (line) density $\mathfrak{n}$, where it can be expressed as
\begin{align}\label{eq:saturation_classical}
    \frac{ds(z)}{dz}=-\frac{\mathfrak{n}\,\sigma(z)}{A}s(z),
\end{align}
with a non-linear scattering cross section
\begin{align}\label{eqe:cross_section}
    \sigma(z)=\frac{\sigma_0}{1+s(z)},
\end{align}
where $\sigma_0=\frac{3\lambda^2}{2\pi}$ is the unsaturated (free space) cross section and $A$ an effective mode area. This is just the semi-classical propagation equation of light propagating through an ensemble of saturable absorbers in forward-scattering approximation~\cite{Veyron_2022, Kim_2020}.

\begin{figure*}[t]
    \centering
    \includegraphics[width=2\columnwidth]{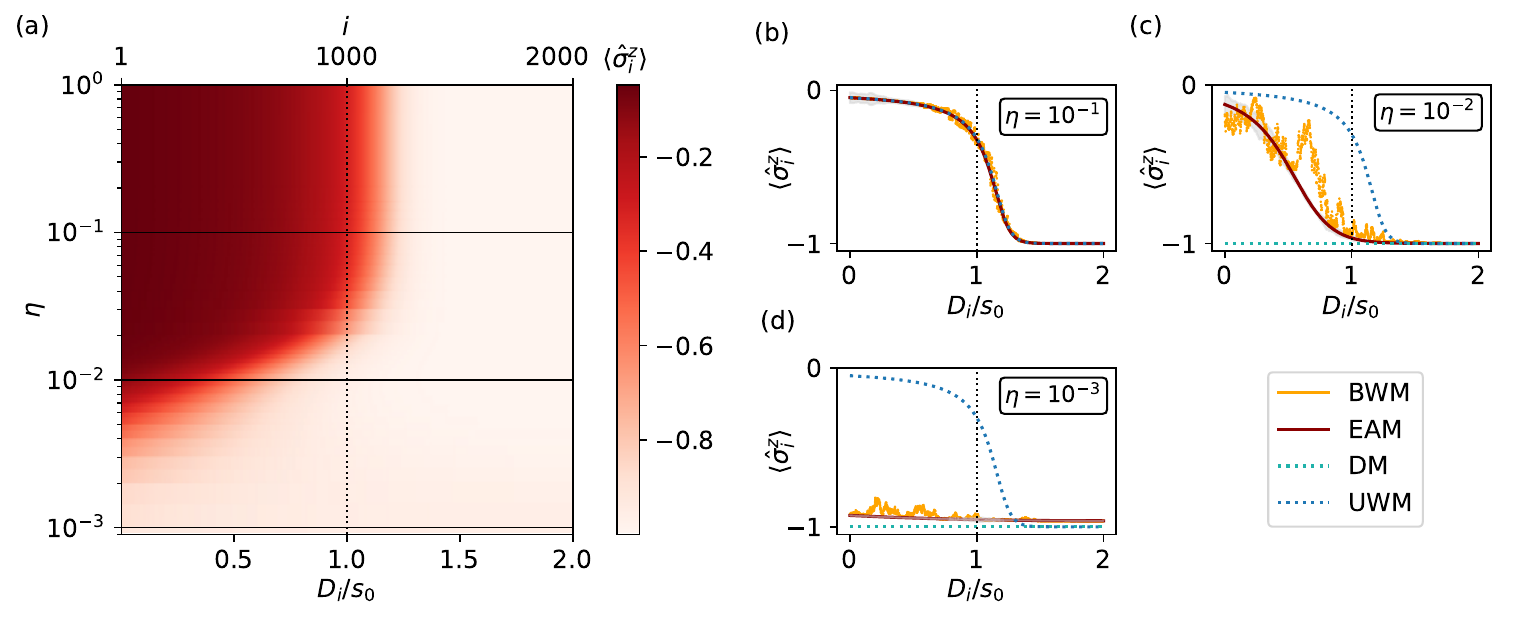}
    \caption{(a) Inversion $\expval{\hat{\sigma}_i^z}$ of an ensemble of bidirectional waveguide systems (EAM) versus $\eta$ and $D_i$ for fixed input saturation $s_0=20$, total emitter number $N=2000$ and coupling strength $\beta=0.005$. For $\eta\geq 0.1$ the dynamics is effectively a unidirectional one. (b)-(d) Cuts of the density plot in (a) at $\eta=0.1, 0.01, 0.001$, compared to results of a single realizations of the system (BWM) and the limits of dissipative Dicke (DM) and unidirectional (UWM)  dynamics. }
    \label{fig:BiWG_sigz}
\end{figure*}

The solution to Eq.~\eqref{eq:saturation_MF} for an initial saturation of the leftmost emitters $s(0)=s_0=2\Omega^2/\Gammatot^2$ is
\begin{align}
    s(D)=\mathcal{W}\left(\tilde{s}De^{(\tilde{s}-1)D}\right),
\end{align}
where $\mathcal{W}(z)$ denotes the Lambert W-function \footnote{The Lambert W-function is the solution to the equation $z=\mathcal{W}(z)e^{\mathcal{W}(z)}$}, and we used the scaled saturation parameter $\tilde{s}=s_0/D$ introduced already in the preceding section. 
We now consider the thermodynamic limit of a large emitter number, or equivalently, large $\OD$ at constant $\tilde{s}$. Using $\mathcal{W}(0) = 0$, one can identify two regimes in the limit $\OD\rightarrow \infty$,
\begin{equation}
    s(\OD) = \begin{cases} 
   0 & \text{for } \tilde{s} <  1 \\
   (\tilde{s}-1)\OD & \text{for } \tilde{s} >  1 
\end{cases}.
\end{equation}
In the first case, the optical depth is large enough such that no light reaches the position $z=\frac{D}{4\beta\mathfrak{n}}$ in the ensemble, while in the second case a large portion of the light still arrives there. Interestingly, the transition between both regimes becomes sharp for large optical depths, as also recently pointed out in \cite{ruostekoski_superradiant_2024,goncalves_driven-dissipative_2024,agarwal2024}, witnessing what has been referred to as a phase separation in~\cite{goncalves_driven-dissipative_2024}. The emergence of a regime of clear phase separation is illustrated in Fig.~\ref{fig:phasesep}(a), which shows the scaled saturation parameter $s(\OD)/s_0$ versus the scaled optical depth $\OD/s_0=1/\tilde{s}$ for various input saturations $s_0$.

This phase separation can also be characterized in terms of the properties of the two-level emitters. The inversion of an emitter, also referred to as magnetization, at position $z(\OD)$ in the chain is $\langle\hat{\sigma}^z(\OD)\rangle = -\frac{1}{1+ s(\OD)}$, as shown in Fig.~\ref{fig:phasesep}(b). In the limit of high input intensity $s_0$, the first emitters along the ensemble are fully saturated, with $\langle\hat{\sigma}^z\rangle= 0$. This drops sharply at the point $\tilde{s} = 1$, and then reaches $\langle\hat{\sigma}^z\rangle= -1$, which corresponds to emitters in their ground state. In this sense, there is a sharp spatial separation between an unpolarized and a fully polarized phase. An order parameter referring to the entire ensemble is given by the mean polarization
\begin{align}
    j_z=\frac{1}{N}\int_0^L\mathrm{d}z n_{\textrm{1D}}\langle\hat{\sigma}^z(z)\rangle
    =\frac{1}{\OD}\int_0^\OD\mathrm{d}\OD' \langle\hat{\sigma}^z(\OD')\rangle,
\end{align}
which is shown in Fig. \ref{fig:phasesep}(c) as a function of $\tilde{s}$ for various total optical depths $\OD$, further illustrating the critical point at $\tilde{s}=1$.

We have established so far all relevant models and characterized the behavior of the system in the limit of complete order and disorder. In the next section, we will proceed by solving the mean-field equations for a range of the disorder parameter $\eta$, analyzing the crossover between these limiting systems and the dependence of the critical value on $\eta$.
\section{Crossover from dissipative Dicke model to unidirectional waveguide model}\label{sec:Numerics}
We are now in a position to explore the crossover from the DM to the UWM based on the mean-field solutions of the EAM, as well as of particular realizations of the BWM assuming fixed random positions between the emitters. We remind the reader that all relevant equations are summarized in Table~\ref{tab:models}. In the following, we will consider the emitter's inversion and the output intensities (saturation parameters) for left and right-propagating fields for a given level of disorder $\eta$ in the EAM and BWM. This allows us to explore a transition from DM to UWM and their respective phases by tuning from $\eta=0$ to $\eta\rightarrow\infty$ (where $\eta\simeq 1$ is sufficient for the current parametrization of disorder).
\subsection{Inversion}
In Fig. \ref{fig:BiWG_sigz} we show $\langle \hat{\sigma}_i^z\rangle$ for $\eta\in [0,1]$ along a chain of $N=2000$ emitters driven with a strength $s_0=20$. We use the same definitions of the intensive input saturation $\tilde{s}$ and optical depth $\ODi$ as in Sec.~\ref{sec:chiral_limit} with the difference that the ensemble does not extend along a continuous axis, but along a discrete lattice.
Figure~\ref{fig:BiWG_sigz}(a) shows the solutions of the EAM exhibiting a phase separation \cite{goncalves_driven-dissipative_2024} around $\tilde{s}=1$ for disorders $\eta\geq0.08$. In the first phase, defined by $1/\tilde{s}<1$, emitters are nearly saturated. In the second phase for $1/\tilde{s}>1$ emitters are in the ground state. More ordered systems show a similar behavior, but earlier in the array. 
For fairly ordered systems, around $\eta\sim 10^{-3}$, emitters are almost completely in the ground state and there is no differentiation of phases.
\begin{figure}[t]
    \centering
    \includegraphics[width=\columnwidth]{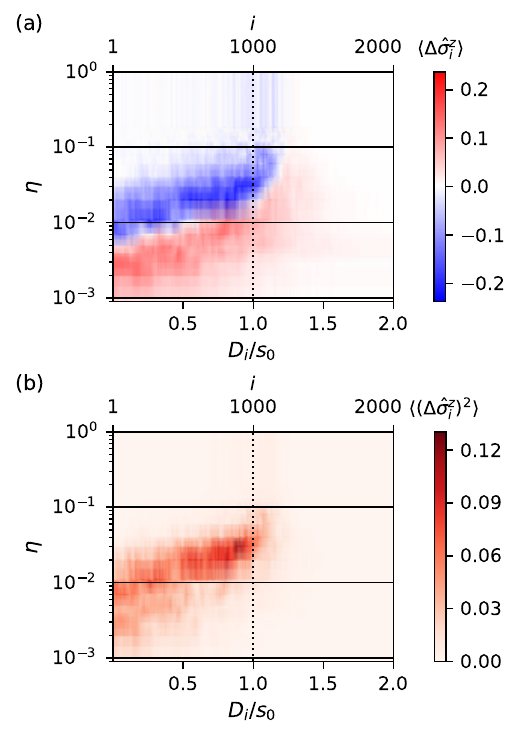}
    \caption{(a) Difference $\expval{\Delta\hat{\sigma}_i^z}$ between the EAM  and 20 realization of the BWM with respect to $\hat{\sigma}_i^z$ for the same set of parameters as in Fig.~\ref{fig:BiWG_sigz}, i.e. total emitter number $N=2000$, fixed input saturation $s_0=20$ and coupling strength $\beta=0.005$. Averaging at the level of solutions to the BWM predicts a phase separations which are not as sharp as in the case of the EAM but with the same critical values. From $\eta=0.1$ on there is no difference between averaging at the level of equation of motions, that is the EAM, and averaging at the level of solutions to these equations.
    (b) variance $\expval{\Delta\hat{\sigma}^z_i}^2$ between the EAM and $20$ realizations of the BWM for the same set of parameters as (a). The variance gets maximal along the border of the phase separation and for disorder values of $\eta\sim0.02$.
    }
    \label{fig:Sigz_Diff&Var}
\end{figure}
Figures \ref{fig:BiWG_sigz}(b)-(d) compare results of the EAM and BWM to the solutions of the UWM and the DM. In the regime of large disorder in Fig. \ref{fig:BiWG_sigz}(b), the BWM, EAM and UWM results agree, highlighting that the random phases the light acquires traveling along the chain effectively cancel for disorders above a threshold.
Interestingly, this disorder does not need to be large. It is for modest amounts of disorder $\eta\sim 0.01$ that fluctuations in the emitter's solutions are the largest, cf. Fig. \ref{fig:BiWG_sigz}(c). In Fig. \ref{fig:BiWG_sigz}(d) letting $\eta\rightarrow 0$, that is approaching the Bragg regime, the fluctuations get again smaller until they vanish completely for $\eta=0$. The dependence of the dynamics on $\eta$ can be understood via the framework of phase matching, showing that we do not only have a changes of phase with respect to $\tilde{s}$, but also with respect to the disorder parameter $\eta$. Above $\eta=0.08$, the acquired phases are random enough to result in destructive interference of the left-propagating light. For modest amounts of disorder, that is $\eta\sim 0.01$, light interferes constructively at some distances and destructively at others. Hence, the increased fluctuations in the inversion. For $\eta\rightarrow 0$, the phase matching condition for constructive interference is fulfilled, resulting in a build-up of a coherent left-propagating light field, as will be seen in the next subsection.

The EAM is obtained by taking the ensemble average at the level of the equations of motion of the BWM. In general, such an averaging procedure is not equivalent to the ensemble average at the level of solutions to the equations of motion, if the observables depend nonlinearly on the averaged variable, as is the case here. The difference between these two averaging procedures can be measured via the difference and variance, e.g. with respect to $\expval{\hat{\sigma}_i^z}$,
\begin{align}
\expval{\Delta\hat{\sigma}^z_i}&=\frac{1}{M}\sum_{\mu=1}^M\qty(\expval{\hat{\sigma}^z_i}_{\mathrm{BWM},\mu}-\expval{\hat{\sigma}^z_i}_\mathrm{EAM})\\
    \expval{(\Delta\hat{\sigma}^z_i)^2}&=\frac{1}{M}\sum_{\mu=1}^M\qty(\expval{\hat{\sigma}^z_i}_{\mathrm{BWM},\mu}-\expval{\hat{\sigma}^z_i}_\mathrm{EAM})^2
\end{align}
where $\expval{\hat{\sigma}^z_i}_{\mathrm{BWM},\mu}$ is the expectation value of the $\mu$--th realization of the BWM of an ensemble size $M$.
In Fig.~\ref{fig:Sigz_Diff&Var}(a) we show the difference between the EAM and $n=20$ realizations of the BWM for the same set of parameters as in Fig.~\ref{fig:BiWG_sigz}. While there is certain deviation between the averaging procedures for an intermediate amount of disorder, the line separating the phases is the same in both averaging procedures. The solutions to the BWM appears to predict less inversion in the ensemble before the separation and more after it. That is, the phase separation in this case is not as sharp as predicted by the EAM. Approaching the limit of no disorder, the difference gets smaller, as in the limit of large disorder, where there is effectively no deviation anymore.

Fig.~\ref{fig:Sigz_Diff&Var}(b) shows the corresponding variance between the averaging procedures. The greatest deviation is in the disorder regime around $\eta\sim 0.02$, whereas for higher disorders, that is in the unidirectional regime, both averaging procedures coincide. Approaching the ordered regime, the variance decreases as the dynamical dependence on the positional distribution of the emitters gets smaller.

\subsection{Output fields}
\begin{figure*}[t]
    \centering
    \includegraphics[width=2.05\columnwidth]{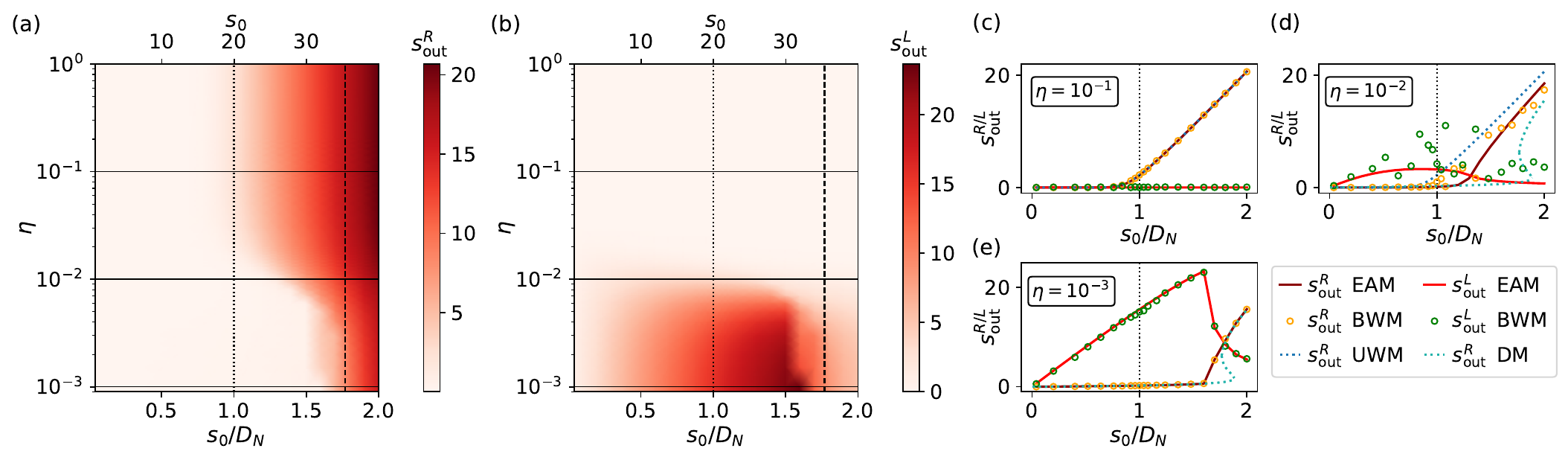}
    \caption{
    (a) Output intensity to the right, as measured by the saturation a hypothetical (N+1)--th emitter at position $z_{N+1}$ would see coming from the left, $s_\mathrm{out}^R$ versus $\tilde{s}$ and $\eta$ for the EAM. The parameters are $N=1000$ and $\beta=0.005$, which corresponds to a total optical depth of $\ODtot=20$. For high $\eta$ the dynamics is effectively a unidirectional one, exhibiting an phase separation at $\tilde{s}=1$. For lower $\eta$ this transition point is shifted to higher input saturation. The dashed black line corresponds to the beginning of the multistable region of the driven-dissipative Dicke model at $s_0= s_-^\mathrm{crit}\simeq 1.77\ODtot$ for the chosen parameters.
    (b) Output intensity to the left, as measured by the saturation a hypothetical 0--th emitter at position $z_{0}$ would see coming from the right, $s_\mathrm{out}^L$ in dependence of $\tilde{s}$ and $\eta$ for the EAM. For disordered emitters there is no coherent left-propagating light field. Around $\eta\sim 0.01$ some coherent left scattering is starting to take place. For $\eta\sim 0.5\cdot10^{-3}$ there is a multistability in the solutions of the nonlinear mean-field equations around $s_0=32$. 
    (c)-(e) Comparison of $s_\mathrm{out}^R$ and $s_\mathrm{out}^L$ calculated for, respectively, single realizations, the ensemble average and the limits of Bragg and unidirectional dynamics.
    }
    \label{fig:BiWG_aR-aL}
\end{figure*}

In Fig.~\ref{fig:BiWG_sigz} we observed the dynamics along the chain at a fixed driving strength. Now, we can take the complementary viewpoint observing the dynamics at specific points of the chain at varying driving strengths. Fig.~\ref{fig:BiWG_aR-aL} displays the coherent output expressed as the saturation a hypothetical ($N+1$)--th emitter would see coming from the left $s_\mathrm{out}^{R}=8\beta\abs*{\langle\hat{a}^R_\mathrm{out}\rangle}^2/\Gamma_\mathrm{tot}$
at the right and the saturation a hypothetical 0--th emitter would see coming from the right $s_\mathrm{out}^{L}=8\beta\abs*{\langle\hat{a}^L_\mathrm{out}\rangle}^2/\Gamma_\mathrm{tot}$ at the left end of a chain of $N=1000$ (corresponding to $\ODtot=20$) emitters for different input saturations $s_0$ and varying disorders $\eta$. Here, the field operators are $\hat{a}^R_\mathrm{out}=\hat{a}(z_N^+,t\rightarrow\infty)-\hat{a}^L_\mathrm{in}$ and $\hat{a}^L_\mathrm{out}=\hat{a}(z_1^-,t\rightarrow\infty)-\hat{a}^R_\mathrm{in}$ for each respective model in steady state. 
In the weak drive regime $s_0/\ODtot<1$ there is essentially no light coming out of the right end of the waveguide, the system is too dense, cf. Fig. \ref{fig:BiWG_aR-aL}(a). This is the case for ordered as well as disordered emitters. Only for higher input power, i.e. $s_0/\ODtot>1$ we see some output building up. More ordered systems exhibit a similar behavior with the difference that the transition point at which $s_\mathrm{out}^R$ is nonzero is achieved at higher driving strengths. The vertical dashed black line indicates the beginning of the bistable regime of the DM (at $s_0= s_-\simeq 1.77\ODtot$ for the chosen parameters and finite system size). The transitions at the critical points are here still fuzzy, since we are not in the thermodynamic limit.
The output saturation $s_\mathrm{out}^L$ in Fig.~\ref{fig:BiWG_aR-aL}(b) shows a different behavior. The disordered chain exhibits no build-up of a coherent left-propagating mode. It is only starting around $\eta\approx 0.01$ that a left-propagating coherent field accumulates and increases with smaller $\eta$ and larger $s_0$. The build-up of the left-propagating field with increasing input saturation breaks down at the same saturation value where the right-propagating field emerges from the ensemble.

In Fig.~\ref{fig:BiWG_aR-aL}(b) and less pronounced also in (a), there is a step-like structure occurring for very low $\eta$, indicating a sudden shift of the input saturation necessary to achieve the break down of $s_\mathrm{out}^L$. This step is due to a multistability of the nonlinear mean-field equations of the EAM, reminiscent of the bistability in the DM. In this region, it is hard to characterize the output field numerically due to the bistability of the system.

In Fig. \ref{fig:BiWG_aR-aL}(c)-(e) we compare saturation intensities at the left and right end of single realizations with the ensemble average and the limiting cases of Bragg and unidirectional dynamics. Regarding single realizations (BWM) each point is calculated for a different distribution of the emitters. We can observe a similar behavior with respect to $\eta$ as in Fig. \ref{fig:BiWG_sigz}. In Fig. \ref{fig:BiWG_aR-aL}(c), for $\eta=0.1$, the dynamics is effectively unidirectional with no coherent left-propagating mode and a phase separation occurring around $\tilde{s}=1$.  
Fig. \ref{fig:BiWG_aR-aL}(d) shows the intermediate regime $\eta=0.01$. The ensemble averaged dynamics is dominated in the first phase ($\tilde{s}<1$) by the left-propagating field and in the second ($\tilde{s}>1$) by the right-propagating one. The fluctuations in the solutions of the BWM are here the largest. 
Finally, for $\eta=0.001$ the dynamics is similar to Bragg physics, cf. Fig \ref{fig:BiWG_aR-aL}(e). As long as $s_0\leq32$ is fulfilled, the array behaves as a mirror and scatters the incoming light completely to the left. Above this point, $s_\mathrm{out}^L$ is breaking down and $s_\mathrm{out}^R$ rises. The comparison with the DM shows almost perfect agreement with the difference that the critical point in the DM is achieved for higher input values. 

\subsection{Correlations between emitters}\label{sec:cumu_exp}

The discussion in the two previous sections clearly illustrates the critical point of $\tilde{s}=1$ in the UWM in terms of the emitter's inversion and output field predicted by mean-field theory. In this section, we consider a higher-order expansion which allows us to study higher-order observables, in particular correlations between emitters and fluctuations in the emitters and in the light field beyond the mean-field approximation, extending and complementing the analysis of~\cite{goncalves_driven-dissipative_2024}. We exploit here the feature of the effective linearity of the UWM discussed in Sec.~\ref{sec:chiral_limit}.

\begin{figure}[t]
	\centering
	\includegraphics[width=\columnwidth]{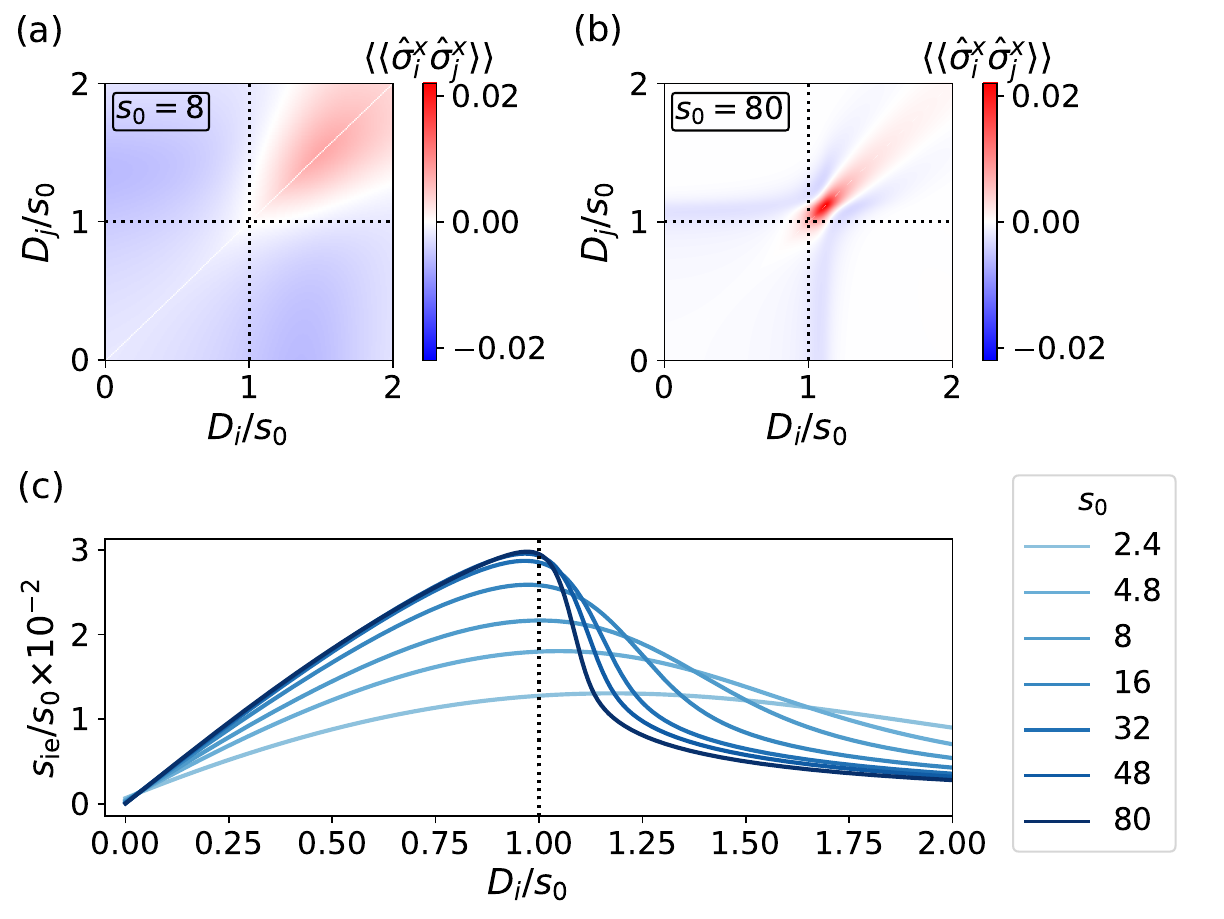}
	\caption{(a),(b) show the second order cumulant $\cexpval{\hat{\sigma}_i^x\hat{\sigma}_j^x}$ for fixed input saturations $s_0=8$ and $s_0=80$, respectively. In (a), even far from the thermodynamic limit, the point $\tilde{s}=1$ is significant, marking a change of behavior in the correlation landscape of the emitters. In (b) a clear phase separation emerged where \mbox{(anti-)correlations} are centered around $\tilde{s}=1$. (c) Inelastic scattered transmission $s_\mathrm{ie}/s_0$ along the chain for several input saturations, reflecting the phase separation for large drive.}
	\label{fig:sigxx&P_inel}
\end{figure}

In Fig.~\ref{fig:sigxx&P_inel}(a) and (b) the second-order cumulant $\cexpval{\hat{\sigma}_i^x\hat{\sigma}_j^x}=\expval{\hat{\sigma}_i^x\hat{\sigma}_j^x}-\expval{\hat{\sigma}_i^x}\expval*{\hat{\sigma}_j^x}$ is shown for the input powers $s_0=8$ and $s_0=80$, respectively. That is, (a) is far away from the thermodynamic limit. First-order observables do not show any signs of a phase separation in this regime of $s_0$ (cf. the line for $s_0=7$ in  Fig. \ref{fig:phasesep}(b) and (c)). However, for second-order observables the point $\ODi/s_0=1$ still seems to have significance, as it marks the transition from anticorrelation to correlation of nearest neighbors. 

Approaching the thermodynamic limit with $s_0=80$ in Fig.~\ref{fig:sigxx&P_inel}(b), we can make two observations: On the one hand, there are no significant correlations among emitters prior to the critical point $\ODi/s_0=1$. Correlations as well as anticorrelations of long- and  short-range nature start to emerge only slightly before that point. On the other hand, the greatest (anti-)correlations occur directly after the critical point and become weaker further along the chain. This behavior is reminiscent of traditional phase transitions, where the greatest fluctuations usually occur at the critical point. Although the thermodynamic limit $\ODi\rightarrow\infty$ is not efficiently computable in second-order cumulant expansion, we assume that the progress we observed from (a) to (b) will hold and, in the end, all \mbox{(anti-)correlations} will involve at least one emitter at the critical point $\OD_i/s_0=1$.

Complementing Fig.~\ref{fig:BiWG_aR-aL}, we show in Fig.~\ref{fig:sigxx&P_inel}(c) the saturation due to inelastically scattered power
\begin{align}
    s_\mathrm{ie}=\frac{8\beta}{\Gammatot}\left(\langle\hat{a}_\mathrm{out}^\dagger\hat{a}_\mathrm{out}\rangle-\abs{\expval{\hat{a}_\mathrm{out}}}^2\right) =8\beta^2\sum_{i,j}^{N}\cexpval{\hat{\sigma}^+_i\hat{\sigma}^-_j}
\end{align}
along the chain normalized to $s_0$ in steady-state.  As light gets scattered along the chain, $s_\mathrm{ie}$ increases and peaks around the critical point $\ODi/s_0$. As with emitter's fluctuations, the peak is attained even far from the thermodynamic limit. However, this does not apply for very low input saturation, e.g. $s_0=2.4$. Since $s_\mathrm{ie}$ is the sum of the integrated in-phase and out-of-phase quadrature fluctuations \cite{Kusmierek2023SCIPOST}, those quantities peak at $\ODi/s_0\approx 1$ as well. 

\subsection{Doppler broadening}
The previous sections and the recent articles \cite{ruostekoski_superradiant_2024,goncalves_driven-dissipative_2024,agarwal2024} discussed large ensembles of identical emitters as can naturally be found in cold atom systems. However, most applications deal with emitters that show additional broadening.
Here, we focus on the case of Doppler broadening, as would be, for example, encountered in gas cells of atomic gases.
Including the detuning $\Delta = \omega - \omega_0$, Eq.~\eqref{eqe:cross_section} is modified into
\begin{equation}
    \sigma(I(z)) = \sigma_0 \frac{1}{1 + s(z) + 4\left(\frac{\Delta_i}{\Gammatot}\right)^2}, \label{eq_sigma_det}
\end{equation}
where $\Delta_i$ is the detuning of the $i$-th emitter. 
Due to the different velocities of each emitter, the detuning follows a Gaussian distribution with standard deviation $\xi_\Delta =\nu_0 \sqrt{k_B T/(m c^2)} $ where $T$ is the temperature of the ensemble, $k_B$ the Boltzmann constant, $m$ the atom's mass, and $c$ the speed of light.
 
\begin{figure}[t]
    \centering
    \includegraphics[width=0.8\linewidth]{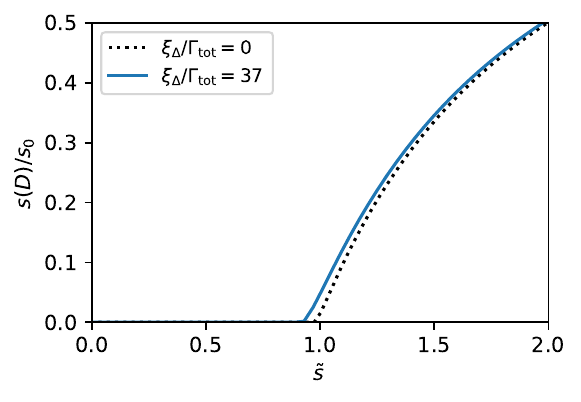}
    \caption{The transmission due to elastic scattering, which is a directly observable quantity as a function of the intensive saturation parameter $\tilde{s} = s_0/\OD$.}
    \label{fig:homogeneous_broadening}
\end{figure}

In Fig.~\ref{fig:homogeneous_broadening} we numerically solve Eq.~\eqref{eq:saturation_classical} with a Gaussian disorder in detuning. Interestingly, the phase separation signatures are still present at the mean-field level while a small smoothing of the kink occurs. The value $\xi_\Delta = 37\Gammatot$ would correspond to the Doppler broadening of a Rubidium gas probed at the D2-line at room temperature. Note that here we kept the definition of $s$ and $\OD$ from the previous section, while neglecting additional effects such as hyperfine structure.

Even though a beyond mean-field simulation including Doppler broadening is out-of-reach for our numerical computation capacity, we would like to point out, that we expect that the results of Sec.~\ref{sec:cumu_exp}, which show that quadrature fluctuations peak at the critical value $\tilde{s} = 1$, would also be valid for a system with Doppler broadening. 
\section{Summary \& outlook} \label{sec:summaryAndOutlook}
We characterized the transition from an ordered to a disordered driven one-dimensional emitter chain with bidirectional interaction in mean-field theory.  We analyzed these limits and the transition between them by means of the emitters' inversion as well as coherently scattered light. In the limit of complete order, realized via a Bragg configuration, the system is described by a driven-dissipative Dicke model exhibiting a well-known first-order phase transition. The opposite case of disorder is effectively described by a unidirectional model.

In this limit, the description of coherent light propagation is essentially the same as semiclassical ad hoc descriptions employed in dealing with ensembles of saturable absorbers. This gives those descriptions a firmer theoretical motivation.
Although this model is not permutationally symmetric, it exhibits a similar feature to the DM, namely a phase separation. In the thermodynamic limit $\OD\rightarrow\infty$ the critical point lies at $\tilde{s}=1$, which is half the value compared to the DM. 
Interestingly, not much disorder is needed for the system to be in the unidirectional regime. This indicates that a larger class of systems, which have some spatial disorder and couple effectively to a one-dimensional bath, can be modeled by the UWM. For example, the light propagation in an elongated cigar-shaped atomic cloud in free space was modeled on different occasions by the DM~\cite{ferioli_non-equilibrium_2023} and UWM in mean-field approximation~\cite{goncalves_driven-dissipative_2024} showing clearly a better and more intuitive description by the latter.
Doppler broadening, which real systems are subject to, does not have any effect on the predictions and validity of the UWM.
Going beyond mean-field theory, using cumulant expansion of order 2, reveals that even far away from the thermodynamic limit the critical point $\tilde{s}=1$ marks a transition in the emitter's correlation landscape, albeit not a phase transition. The saturation resulting from inelastic scattering peaks around this point and with that the integrated in-phase and out-of-phase quadrature fluctuations. 

Our work can be extended in several ways.  
Firstly, including higher-order correlations in the description of the EAM and BWM, as was done for the UWM, could provide more insight into the dependence of light fluctuation and incoherently scattered light on the degree of order $\eta$. This would also make it possible to study the change in correlation lengths at the critical point with varying disorder.  Of course, the higher computational complexity will limit the simulatable system sizes to smaller ones.
Secondly, to model state-of-the-art experiments as \cite{ferioli_non-equilibrium_2023} one can use Green's tensor in free space, as was done in \cite{agarwal2024}, but again with variable degree of order.
Thirdly, the effect of time delays can be incorporated in the theoretical description. In \cite{Sinha2020, Windt2024} it was shown that the properties of superradiance and effective decay rates in one-dimensional waveguide systems are highly influenced in the presence of time delay. Finally, disorder appears to play a key role in shifting from the DM to the UWM dynamics. It is well known that one-dimensional disordered systems possess Anderson localization \cite{Evers2008RMP} and one would expect such modes to also appear in the disordered ensemble. Determining whether these play a key role in the DM to UAM transition is an outstanding question.
\\\\
\begin{acknowledgments}
We are grateful to S. Felicetti, C. Lidl, A. Rauschenbeutel, T. Roscilde, P. Schneeweiss \& J. Volz for stimulating
discussions. 
This work was funded by the Deutsche Forschungsgemeinschaft (DFG, German Research Foundation) under SFB 1227 ‘DQ-mat’ project A09 - 274200144 and Germany’s Excellence Strategy – EXC-2123 QuantumFrontiers – 390837967. S.M. acknowledges support from the Australian Research Council (ARC) via the Future Fellowship, `Emergent many-body phenomena in engineered quantum optical systems', project no. FT200100844. M.S. acknowledges support via the Italian Ministry of University and
Research, in the framework of the National Recovery and Resilience Plan (NRRP) for funding project MSCA 0000048. 
\end{acknowledgments}

\sloppy
\bibliography{bigBib, phase_transition}

\appendix

\section{Master equation}
\label{sec:app_MEQ}

Here we provide a derivation of the bidirectional master equation, presented in (\ref{eq:Bi-MEQ}). We begin with a general master equation for an ensemble of two-level emitters with resonance frequency $\omega_0$ coupled to an arbitrary electromagnetic environment \cite{Dung2002PRA, Asenjo2017PRA, Asenjo2017PRX} where each emitter is driven by a coherent field,
\begin{equation}
\frac{d \hat{\rho}}{d t} = -i \left[\hat{H}_{\rm eff}, \hat{\rho} \right] + \sum_{i,j=1}^N \frac{\Gamma_{i j}}{2} \left[2 \hat{\sigma}^-_i \hat{\rho} \hat{\sigma}^+_j - \hat{\sigma}_i^+ \hat{\sigma}_j^- \hat{\rho} - \hat{\rho} \hat{\sigma}_i^+ \hat{\sigma}_j^- \right],
\end{equation}
with 
\begin{equation}
\begin{split}
\hat{H}_{\rm eff} &=  -\sum_{i=1}^N \Delta \hat{\sigma}^+_i \hat{\sigma}^-_i + \frac{\Omega_i}{2} \hat{\sigma}_i^- + \frac{\Omega_i^*}{2} \hat{\sigma}_i^+ +\sum_{i,j=1}^N J_{ij} \hat{\sigma}_i^+ \hat{\sigma}_j^-,\\
J_{i j} &= - \mu_0 \omega_0^2 \, \mathbf d^* \cdot \operatorname{Re}{\left[ \mathbf{G}(\mathbf r_i, \mathbf r_j, \omega_0) \right]} \cdot \mathbf d , \\
\Gamma_{i j} &=  2 \mu_0 \omega_0^2 \, \mathbf d^* \cdot \operatorname{Im}{\left[ \mathbf{G}(\mathbf r_i, \mathbf r_j, \omega_0) \right]} \cdot \mathbf d.
\end{split}.
\end{equation}

%

Here $\mathbf{G}(\mathbf r, \mathbf r', \omega_0)$ is Green's tensor for the electromagnetic environment evaluated at the resonance frequency of the emitters. By taking only Green's tensor at the resonance frequency of the emitters, non-Markovian effects of the electromagnetic environment and time delays in the emitter interactions are neglected. This requires that we have $L \ll v_g/\Gammatot$ where $v_g$ is the group velocity of light inside the waveguide, and $L$ the length of the ensemble, and as we explain further below, $\Gammatot$ is the total decay rate of the emitter. This condition is typically fulfilled well in quantum optics experiments. Furthermore, interactions due to Casimir effects are also neglected \cite{Asenjo2017PRX}. Under these approximations, the positive frequency component electric field operator is \cite{Dung2002PRA, Asenjo2017PRA, Asenjo2017PRX},

\begin{align}
\label{eq:electricFieldOperator}
\hat{\mathbf{E}}^{+}(\mathbf{r})=\hat{\mathbf{E}}_{\mathrm{in}}^{+}(\mathbf{r})+\mu_0 \omega_0^2 \sum_{j=1}^N \mathbf{G}\left(\mathbf{r}, \mathbf{r}_j, \omega_0\right) \cdot \mathbf d \,\hat{\sigma}^{-}_j.
\end{align}
Here, $\hat{\mathbf{E}}^{+}(\mathbf{r})$ ($\hat{\mathbf{E}}^{-}(\mathbf{r})$) is the positive frequency (negative frequency) component field operator for the total field and $\hat{\mathbf{E}}_{\mathrm{in}}^{+}(\mathbf{r})$ ($\hat{\mathbf{E}}_{\mathrm{in}}^{-}(\mathbf{r})$) is the positive frequency (negative frequency) component field operator for the input field. We note that, for brevity, the dependence of the operators on time is not shown.

To proceed, we now make two assumptions about the nature of the ensemble. (i) The ensemble is arranged in a line or as a pencil shape such that the optical depth of the ensemble is negligible in the transverse direction and the 3D~\cite{sorensen_three-dimensional_2008} problem becomes effectively 1D. These assumptions are appropriate when considering experimental geometries such as cold atoms coupled to nanofibers. (ii) We assume that the ensemble is dilute such that its number density is less than one emitter per cubic wavelength. This will allow us to neglect collective effects due to emission into the non-guided modes of the environment.

Green's tensor can be expanded in terms of the normal modes of the electromagnetic environment \cite{Novotny2012Book, Lodahl2015RMP},
\begin{align}
\label{eq:greensTensor}
\mathbf{G}(\mathbf r, \mathbf r',\omega_0) = c^2 \int d\mathbf k \frac{\mathbf E_\mathbf k(\mathbf r) \mathbf E^*_\mathbf k(\mathbf r')}{\omega_\mathbf k^2 - \omega_0^2}.    
\end{align}
Here, the integral implies summation over all electromagnetic modes of the system, i.e. both discrete and continuous. For an ensemble coupled to a waveguide, it is then possible to write Green's tensor as a sum of contributions for a subset of normal modes that form a one-dimensional continuum of waveguide modes $\mathbf{G}_{\rm 1D}$, and a contribution from all other modes $\mathbf{G}_{0}$. We can thus write
\begin{equation}
    \mathbf{G}(\mathbf r, \mathbf r',\omega_0) = \mathbf{G}_{\rm 1D}(\mathbf r, \mathbf r',\omega_0) + \mathbf{G}_{0}(\mathbf r, \mathbf r',\omega_0).
\end{equation}
The cooperative decay rates and frequency shifts therefore become
\begin{equation}
\begin{split}
J_{i j} &= - \mu_0 \omega_0^2 \, \mathbf d^* \cdot \operatorname{Re}{\left[ \mathbf{G}_{\rm 1D}(\mathbf r_i, \mathbf r_j,\omega_0) + \mathbf{G}_{0}(\mathbf r_i, \mathbf r_j,\omega_0) \right]} \cdot \mathbf d , \\
\Gamma_{i j} &=  2 \mu_0 \omega_0^2 \, \mathbf d^* \cdot \operatorname{Im}{\left[ \mathbf{G}_{\rm 1D}(\mathbf r_i, \mathbf r_j,\omega_0) + \mathbf{G}_{0}(\mathbf r_i, \mathbf r_j,\omega_0) \right]} \cdot \mathbf d.    
\end{split}    
\end{equation}
Using assumptions (i) and (ii), we can neglect collective coupling terms and collective decay terms that occur through modes outside of the one-dimensional continuum. We therefore have
\begin{equation}
\begin{split}
\label{eq:MEratesWithApprox}
J_{i j} &= - \mu_0 \omega_0^2 \, \mathbf d^* \cdot \operatorname{Re}{\left[ \mathbf{G}_{\rm 1D}(\mathbf r_i, \mathbf r_j,\omega_0) \right]} \cdot \mathbf d + \Delta_{ii} \delta_{ij}, \\   
\Gamma_{i j} &=  2 \mu_0 \omega_0^2 \, \mathbf{d}^* \cdot \operatorname{Im}{\left[ \mathbf{G}_{\rm 1D}(\mathbf{r}_i, \mathbf{r}_{j},\omega_0) \right]} 
\end{split}    
\end{equation}
Here $\delta_{ij}$ is the Kronecker delta (returns $1$ when $i=j$ and zero otherwise) and $\Delta_{ii} = - \mu_0 \omega_0^2 \, \mathbf d^* \cdot \operatorname{Re}{\left[ \mathbf{G}_{0}(\mathbf r_i, \mathbf r_i,\omega_0) \right]} \cdot \mathbf d$ is the self shift or Lamb shift of each emitter due to the electromagnetic vacuum. This term can be absorbed into the definition of the emitter resonance frequency $\omega_0$. We also have $\gamma_{ii} = 2 \mu_0 \omega_0^2 \, \mathbf d^* \cdot \operatorname{Im}{\left[ \mathbf{G}_{0}(\mathbf r_i, \mathbf r_i,\omega_0) \right]} \cdot \mathbf d$, which quantifies spontaneous emission of each emitter to the non-guided modes outside of the waveguide. In the example of a nanofiber geometry, the presence of the fiber does not have a significant influence on the emission rate and one can approximate this to be equal to the vacuum spontaneous emission rate of the emitter's transition.

We now assume that the electromagnetic environment that makes up Green's tensor has self-frequency shifts and decay rates that are independent of position in the direction perpendicular to the array. For a nanofiber, this occurs when emitters are coupled to the waveguide such that the radial distance of all atoms from the nanofiber is equal. If the ensemble is coupled to a more complex one-dimensional structure such as a photonic crystal waveguide the assumptions become more stringent: all emitters must be of equal distance from the photonic crystal waveguide and most occupy the same position within a unit cell. Under these constraints, we can thus write $\Delta_{ii} = \Delta$ and $\gamma_{ii} =\gamma$.

We now turn to the one-dimensional part of Green's tensor. A number of works have previously derived Green's tensor for photonic crystal waveguides \cite{MangaRao2007PRB, MangaRao2007PRL} and waveguides with continuous translational symmetry \cite{Asenjo2017PRX}. If Green's tensor is evaluated at a fixed distance from the waveguide center, it has the general form,
\begin{equation}
\begin{split}
    \mathbf{G}_{\rm 1D} (\mathbf r, \mathbf r', \omega_0) = i \mathbf{F}  e^{i k_0 |z - z'|},
\end{split}    
\end{equation}
where we have taken $z$ to be the propagation direction along the waveguide. For a waveguide with continuous translational symmetry, we have the tensor $\mathbf{F} =  \pi c^2/(v_g \omega_0) \mathbf E \mathbf E^T$. Here, $v_g$ is the group velocity of the waveguide modes at the resonance frequency of the emitters and $\mathbf E_{k_0}(\mathbf r) = \mathbf E e^{i k_0 z}$ are the normal modes of the guided waveguide modes evaluated at the wavenumber $k_0$ of the one-dimensional mode corresponding to the emitter's resonance. The position dependence along the cross section of the waveguide has been neglected. We take the electric field of the guided modes $\mathbf E$ to be real valued.

Substituting these rates in the master equation, we obtain
\begin{equation}
\begin{split}
&\frac{d \hat{\rho}}{d t} = -i \left[\hat{H}_{\rm eff}, \hat{\rho} \right] + \sum_{i=1}^N  \frac{\gamma}{2} \left[2 \hat{\sigma}_i^- \hat{\rho} \hat{\sigma}_i^+ - \hat{\sigma}^+_i \hat{\sigma}^-_i \hat{\rho} - \hat{\rho} \hat{\sigma}^+_i \hat{\sigma}^-_i \right]\\
+& \sum_{i,j=1}^N \frac{\Gamma_{\rm 1D}}{2} \cos{\left( k_0 |z_i - z_j| \right)} \left[2 \hat{\sigma}_i^- \hat{\rho} \hat{\sigma}_j^+ - \hat{\sigma}^+_i \hat{\sigma}^-_j \hat{\rho} - \hat{\rho} \hat{\sigma}^+_i \hat{\sigma}^-_j \right],
\end{split}
\end{equation}
with 
\begin{equation}
\begin{split}
\hat{H}_{\rm eff} &=  -\sum_{i=1}^N \Delta \hat{\sigma}_i^+ \hat{\sigma}_i^- + \frac{\Omega_i}{2} \hat{\sigma}^-_i + \frac{\Omega_i^*}{2} \hat{\sigma}^+_i \\
&+ \sum_{i,j=1}^N \frac{\Gamma_{\rm 1D}}{2} \sin{\left( k_0 |z_i - z_j| \right)} \hat{\sigma}^+_i \hat{\sigma}^-_j,
\end{split}
\end{equation}
where $\Gamma_{\rm 1D} = 2 \mu_0 \omega_0^2 \, \mathbf d^* \cdot \mathbf{F} \cdot \mathbf d$. 
\\\\
\section{Field operators, Input-output relations and Rabi frequency} \label{sec:app_inputOutput}

Here we provide details of our derivation of the field operators, input-output relations, and Rabi frequency. Starting from the expression for the field operators, we use the normal modes of the system to expand the electric field and we explicitly separate the expansion to an integral over the one-dimensional waveguide modes and other modes,
\begin{equation}
\begin{split}
    \hat{\mathbf E}^{+}(\mathbf r, t) &= \int_{\rm 1D} dk \sqrt{\frac{\omega_k}{2 \epsilon_0}} \mathbf E_k(\mathbf r) \hat{a}_k(t)\\
    &+ \int d \mathbf k \sqrt{\frac{\omega_k}{2 \epsilon_0}} \mathbf E_\mathbf k(\mathbf r) \hat{a}_\mathbf k(t).
\end{split}
\end{equation}
Equivalently for the input field operator,
\begin{equation}
\begin{split}
    \hat{\mathbf E}_{\rm in}^{+}(\mathbf r, t) &= \int_{\rm 1D} dk \sqrt{\frac{\omega_k}{2 \epsilon_0}} \mathbf E_k(\mathbf r) \hat{a}_k(0) e^{- i (\omega_k - \omega_0)t}\\
    &+ \int d \mathbf k \sqrt{\frac{\omega_\mathbf k}{2 \epsilon_0}} \mathbf E_\mathbf k(\mathbf r) \hat{a}_\mathbf k(0) e^{- i (\omega_\mathbf k - \omega_0)t}.  
\end{split}
\end{equation}
We also do this explicitly for Green's tensor,
\begin{equation}
\begin{split}
\mathbf{G}(\mathbf r, \mathbf r',\omega_0) &= c^2 \int_{\rm 1D} d k \frac{\mathbf E_k(\mathbf r) \mathbf E^*_k(\mathbf r')}{\omega_k^2 - \omega_0^2} \\
&+ c^2 \int d\mathbf k \frac{\mathbf E_\mathbf k(\mathbf r) \mathbf E^*_\mathbf k(\mathbf r')}{\omega_\mathbf k^2 - \omega_0^2}.    
\end{split}
\end{equation}
The electric field of the modes satisfy the orthogonality relation,
\begin{equation}
    \int d \mathbf r \epsilon(\mathbf r) \mathbf E_{\mathbf k}^*(\mathbf r) \cdot \mathbf E_{\mathbf k'}(\mathbf r) = \delta(\mathbf k - \mathbf k'),
\end{equation}
where $\epsilon(\mathbf r)$ is the permittivity distribution of the waveguide and we note that the 1D modes are orthogonal to other unguided modes. Substituting the field expansions into the electric-field operator (\ref{eq:electricFieldOperator}) and using the orthogonality condition to project onto modes of the waveguide, we obtain an expression for the annihilation operator of the waveguide mode with wavenumber $k$,
\begin{equation}
\begin{split}
\label{eq:WGannihilationK}
    &\sqrt{\frac{\omega_k}{2\epsilon_0}} \hat{a}_k(t) = \sqrt{\frac{\omega_k}{2\epsilon_0}} \hat{a}_k(0) e^{- i (\omega_k - \omega_0)t}\\
    &+ \mu_0 \omega_0^2 c^2 \sum_{i=1}^N \frac{\mathbf E_k^*(\mathbf r_i) \cdot \mathbf d}{\omega_k^2 - \omega_0^2}  \hat{\sigma}_i^-(t).
\end{split}    
\end{equation}

We now want to transform this equation to contain real-space creation and annihilation operators for the waveguide modes. When doing this we separate the operators into independent operators for forwards and backwards propagating operators. These can be approximated to commute. This approximation is valid as the resonance frequency $\omega_0$ is much larger than the emitter's bandwidth $\Gamma_{\rm tot}$. Across this bandwidth the dispersion can be assumed to be linear with group velocity $v_g$, i.e., $\omega_k = \omega_0 + v_g (k - k_0)$ and therefore the wavenumbers for forwards and backwards propagating are well separated in $k$-space. To proceed we multiply (\ref{eq:WGannihilationK}) by $e^{i k  z}/\sqrt{2\pi}$ and integrate over all positive wavenumbers. We obtain for the LHS,
\begin{equation}
\begin{split}
    \int_{>0} dk \sqrt{\frac{\omega_k}{4 \pi \epsilon_0}} \hat{a}_k(t) e^{i k  z} &\sim \sqrt{\frac{\omega_0}{4 \pi \epsilon_0}} e^{i k_0 z} \int dk  \hat{a}_k(t) e^{i (k-k_0) z}\\
    &\equiv \sqrt{\frac{\omega_0}{2 \epsilon_0}} \hat{a}^R(z,t).
\end{split}
\end{equation}
Similarly for the input field we have,
\begin{equation}
\begin{split}
    &\int_{>0} dk \sqrt{\frac{\omega_k}{4 \pi \epsilon_0}} \hat{a}_k(0)  e^{-i (\omega_k - \omega_0)t} e^{i k  z}\\
    &\sim \sqrt{\frac{\omega_0}{4 \pi \epsilon_0}} e^{i k_0 z}\int dk \, \hat{a}_k(0) e^{i (k - k_0) (z - v_g t)}\\
    &\equiv \sqrt{\frac{\omega_0}{2 \epsilon_0}} \hat{a}^R(z - v_g t, 0).
\end{split}
\end{equation}
And for the emitter term,
\begin{equation}
\begin{split}
\label{eq:atomicTermInputOutput}
    &\frac{\mu_0 \omega_0^2 c^2}{\sqrt{2\pi}} \sum_{i=1}^N \int_{>0} dk e^{i k z} \frac{\mathbf E_k^*(\mathbf r_i) \cdot \mathbf d}{\omega_k^2 - \omega_0^2}  \hat{\sigma}_i^-(t)\\
    &\sim \frac{\mu_0 \omega_0 c^2}{2 v_g \sqrt{2\pi}} e^{i k_0 z} \sum_{i=1}^N \mathbf E_{k_0}^*(\mathbf r_i) \cdot \mathbf d \hat{\sigma}_i^-(t) \int dk  \frac{e^{i (k - k_0)(z - z_i)}}{k - k_0}\\
    &= \frac{\mu_0 \omega_0 c^2}{2 v_g \sqrt{2\pi}} e^{i k_0 z} \sum_{i=1}^N \mathbf E_{k_0}^*(\mathbf r_i) \cdot \mathbf d \hat{\sigma}_i^-(t) 2 i \pi \theta(z-z_i),
\end{split}
\end{equation}
where $\theta(z)$ is the Heaviside step function and we used the convention $\theta(0)=\frac{1}{2}$.
Here we have used the approximation $\mathbf E_k(\mathbf r) \sim \mathbf E_{k_0}(\mathbf r) e^{i (k - k_0) z}$. We can express the dipole moment as $\mathbf d = \mathbf e_d |\mathbf d|$, where $\mathbf e_d$ is the unit vector pointing in the direction of the transition dipole moment. To obtain the usual input-output relations we want to relate these terms to the emission rate of the emitters to the waveguide $\Gamma_{\rm 1D}$. To do this, we use the definition of the local density of states of a 1D waveguide \cite{Lodahl2015RMP},
\begin{equation}
    \rho(\omega, \mathbf r) = \int d k |\mathbf e_d \cdot \mathbf E_k^*(\mathbf r)|^2\delta(\omega - \omega_k) \sim 2 \frac{|\mathbf e_d \cdot \mathbf E_k^*(\mathbf r)|^2}{v_g}.
\end{equation}
To approximate the second equality we have used the linear dispersion relation approximation, and the factor of 2 comes from the presence of a forwards and backwards propagating mode. The emission rate into the one-dimensional waveguide mode is given by $\Gamma_{\rm 1D} = \pi |\mathbf d|^2 \mu_0 c^2 \omega_0 \rho(\omega_0)$. Combining these, (\ref{eq:atomicTermInputOutput}) is written as,
\begin{equation}
i \sqrt{\frac{\omega_0}{2\epsilon_0}}  \sqrt{\frac{\Gamma_{\rm 1D}}{2 v_g}} \sum_{i=1}^N \hat{\sigma}^-_i(t) \theta(z - z_i) e^{i k_0 (z - z_i)} e^{i \operatorname{arg}{(g_i)}}.
\end{equation}
Here we have defined $g_i = \mathbf E^* \cdot \mathbf e_d$, where $\mathbf E_{k_0}(\mathbf r) = \mathbf E e^{i k_0 z}$, such that $\mathbf E$ does not contain a propagation phase. This means that $g_i$ has the same phase for all emitters. We take this phase to be $\pi$. We then have the relation for the forward-propagating photon annihilation operator,
\begin{equation}
\label{eq:photonFieldForward}
    \hat{a}^R(z,t) = \hat{a}^R(z - v_g t, 0) - i \sqrt{\frac{\Gamma_{\rm 1D}}{2 v_g}} \sum_{i=1}^N \hat{\sigma}^-_i\theta(z-z_i)e ^{i k_0 (z-z_i)}.
\end{equation}
To get the input-output relations we evaluate this equation just after the $N$th emitter, i.e. at $z=z_N^+$. We then define $\hat{a}^R_{\rm out}(t) = \hat{a}^R(z_N^+,t) $ and $\hat{a}^R_{\rm in}(t) = \hat{a}^R(z_N^+ - v_g t,0)$. This gives (\ref{eq:inputOutputRelations1}) in the main text. Following the same process for the backwards propagating modes we obtain an equivalent equation,
\begin{equation}
\label{eq:photonFieldBackward}
        \hat{a}^L(z,t) = \hat{a}^L(z + v_g t, 0) - i \sqrt{\frac{\Gamma_{\rm 1D}}{2 v_g}} \sum_{i=1}^N \hat{\sigma}^-_i\theta(z_i-z)e ^{i k_0 (z_i-z)}.
\end{equation}
Evaluating the field just before the first emitter, i.e. at $z=z_1^-$ and defining $\hat{a}^L_{\rm out}(t) = \hat{a}^L(z_1^-,t) $ and $\hat{a}^L_{\rm in}(t) = \hat{a}^L(z_1^+ + v_g t,0)$, we obtain (\ref{eq:inputOutputRelations2}) in the main text.

To obtain a photon field operator for the entire field we define $\hat{a}(z, t) = \hat{a}^R(z,t) + \hat{a}^L(z,t)$. Combining (\ref{eq:photonFieldForward}) and (\ref{eq:photonFieldBackward}) gives
\begin{equation}
\begin{split}
\hat{a}(z, t) =&   \hat{a}^R(z - v_g t, 0) + \hat{a}^L(z + v_g t, 0)\\
&- i \sqrt{\frac{\Gamma_{\rm 1D}}{2 v_g}} \sum_{i=1}^N \hat{\sigma}^-_ie ^{i k_0 |z - z_i|} .
\end{split}
\end{equation}

We now derive an expression for the Rabi frequency in terms of the input field. We assume the ensemble is driven only by a right-propagating field and that all other fields are in vacuum. Using the same definitions and approximations as we made to derive the input-output relations, we have
\begin{equation}
\begin{split}
    \frac{\Omega_i}{2} &= - \langle \hat{\mathbf E}_{\rm in}^- (\mathbf r_i, t) \cdot \mathbf d \rangle\\
    &= -  \int_{>0} dk \sqrt{\frac{\omega_k}{2\epsilon_0}}  \langle\hat{a}^\dagger_k(0)\rangle e^{ i (\omega_k - \omega_0)t} \mathbf E^*_k(\mathbf r_i) \cdot \mathbf d \\
    &\sim -  c\sqrt{\frac{\omega_0 \mu_0 }{2}} |\mathbf d|\mathbf E_{k_0}^*(\mathbf r_i) \cdot \mathbf e_d \, e^{-i k_0 z_i}\\
    &\times \int dk   \langle\hat{a}^\dagger_k(0)\rangle e^{- i (k - k_0)(z_i - v_g  t)} \\
    &= -  c\sqrt{\omega_0 \mu_0 \pi} |\mathbf d|\mathbf E_{k_0}^*(\mathbf r_i) \cdot \mathbf e_d \langle\hat{a}^{R \, \dagger}(z_i - v_g t, 0)\rangle\\
    &= \sqrt{\frac{\Gamma_{\rm 1D} v_g}{2}} \langle \hat{a}_{\rm in}^{R \, \dagger} \rangle e^{-i k_0 z_i}.
\end{split}
\end{equation}
In the last line, we have assumed that the system is driven by a monochromatic field.

\section{Ensemble average of bidirectional waveguide system}\label{app:ensemble_average}
In this section we take the ensemble average of the BWM. Considering a bidirectional waveguide system described by \eqref{eq:Bi-MEQ} we assume the emitters to be $\lambda/2$ apart and vary their distance according to a Gaussian distribution of the form 
\begin{align}\label{eq:gaussian_app}
    p(Z_j)=\sqrt{\frac{2}{\pi\eta^2\lambda^2}}\exp[-\frac{(Z_j - \frac{\lambda}{2})^2}{2\qty(\eta\frac{\lambda}{2})^2}]
\end{align}
with $Z_j=z_{j+1}-z_j$. On the one hand, this distribution describes the distances between emitters along the chain in one realization of the experiment. On the other hand, it describes the variation in distance of the same pair of emitters in an ensemble of experiments. Before taking the average over the emitter distances, we apply a local transformation $\hat{\sigma}_i^-\rightarrow\hat{\sigma}_i^- e^{ik_0z_i}$ to \eqref{eq:Bi-MEQ} to absorb the driving phase. The master equation in this new local gauge reads
\begin{align}\label{eq:MEQ_spiral}
    \frac{d\hat{\rho}}{dt}=&-i\qty[\hat{H}_\mathrm{sys}+\hat{H}_L+\hat{H}_R,\hat{\rho}]\nonumber\\
    &+ \sum_{ij}\Gamma'_{ij}\qty(\hat{\sigma}^-_i\hat{\rho}\hat{\sigma}^+_j - \frac{1}{2}\hat{\sigma}^+_{i}\hat{\sigma}^-_{j}\hat{\rho} - \frac{1}{2}\hat{\rho}\hat{\sigma}^+_{i}\hat{\sigma}^-_{j})
\end{align}
with 
\begin{align}
    \hat{H}_\mathrm{sys} & = \sum_j\frac{\Omega}{2}\qty( \hat{\sigma}^-_j + \hat{\sigma}_j^{+}) \\
    \hat{H}_R & =-\frac{i  \Gamma_\mathrm{1D}}{4} \sum_{j>l}\left(\hat{\sigma}_j^{+} \hat{\sigma}_l^--\mathrm{ h.c. }\right) \\
    \hat{H}_L & =-\frac{i  \Gamma_\mathrm{1D}}{4} \sum_{j<l}\left(e^{-2ik_0(z_j-z_l)} \hat{\sigma}_j^{+}\hat{\sigma}_l^- -\mathrm{ h.c. }\right) 
\end{align}
and 
\begin{align}
    \Gamma'_{ij}= \gamma\delta_{ij} + \frac{\Gamma_\mathrm{1D}}{2}\qty(1+e^{-2ik_0(z_i-z_j)}).
\end{align}
The spatially varying field transforms as $\hat{a}(z,t)\rightarrow e^{ik_0z}\hat{a}(z,t)$ and reads  
\begin{align}\label{eq:spiral_in-out}
\hat{a}^\mathrm{BWM}(z,t)=&\hat{a}_{\mathrm{in}}(t)-i \sqrt{\frac{\Gamma_{1 \mathrm{D}}}{2}}\sum_{j: z_j<z} \hat{\sigma}_j^{-}(t) \nonumber\\ 
& -i \sqrt{\frac{\Gamma_{1 \mathrm{D}}}{2}} \sum_{j: z_j>z} e^{-2ik_0(z-z_j)} \hat{\sigma}_j^{-}(t).
\end{align} 
In mean-field approximation we get equations $\eqref{eq:MF}$ with the effective Rabi frequency 
\begin{align}\label{eq:effectiv_drive_Pos}
\alpha_i^\mathrm{BWM} =& \frac{\Omega}{2} -i\frac{\Gamma_\mathrm{1D}}{2}\sum_{j=1}^{i-1}\expval*{\hat{\sigma}_j^-}\nonumber\\
&-i\frac{\Gamma_\mathrm{1D}}{2}\sum_{j=i+1}^{N}e^{2ik_0(z_j-z_i)}\expval*{\hat{\sigma}_j^-},
\end{align}
Interested in the average dynamics of the ensemble we take the master equation \eqref{eq:MEQ_spiral} and average over the emitter distances, that is phases, with respect to \eqref{eq:gaussian_app}. The averaged phases are effectively anisotropic distance-dependent scattering rates. Traveling from emitter $j+1$ to $j$ the emitters scatter photons with an additional rate
\begin{align}
    \int_{-\infty}^\infty \dd Z_j p(Z_j) e^{\pm2ik_0 Z_j}=e^{-2(\eta\pi)^2}.
\end{align}
Similarly, the light traveling from emitter $j$ to $i$ is additional scattered with a rate $e^{-2(\eta\pi)^2\abs{i-j}}$. The master equation for the average emitter's dynamics takes the form
\begin{align}
    \frac{d\hat{\rho}}{dt}=&-i\qty[\hat{H}_\mathrm{sys}+\hat{H}_L^\mathrm{av}+\hat{H}_R,\hat{\rho}]\nonumber\\
    &+ \sum_{ij}\Gamma^\mathrm{av}_{ij}\qty(\hat{\sigma}^-_i\hat{\rho}\hat{\sigma}^+_j - \frac{1}{2}\hat{\sigma}^+_{i}\hat{\sigma}^-_{j}\hat{\rho} - \frac{1}{2}\hat{\rho}\hat{\sigma}^+_{i}\hat{\sigma}^-_{j})
\end{align}
with
\begin{align}
    \hat{H}_L^\mathrm{av} & =-\frac{i \Gamma_\mathrm{1D}}{4} \sum_{j<i}e^{-2(\eta\pi)^2\abs{i-j}}\left( \hat{\sigma}_j^{+}\hat{\sigma}_i^--\mathrm{ h.c. }\right) \\
    \Gamma^\mathrm{av}_{ij}&= \gamma\delta_{ij} + \frac{\Gamma_\mathrm{1D}}{2}\qty(1+e^{-2(\eta\pi)^2\abs{i-j}}).
\end{align}
Since the phases only enter in describing the left-propagating mode the corresponding field takes the form 
\begin{align}
\hat{a}^\mathrm{EAM}(z_i,t)=&\hat{a}_{\mathrm{in}}(t)-i \sqrt{\frac{\Gamma_{1 \mathrm{D}}}{2}}\sum_{j: z_j<z_i} \hat{\sigma}_j^{-}(t) \nonumber\\ &-i\sqrt{\frac{\Gamma_{1 \mathrm{D}}}{2}}\sum_{j: z_j>z_i} e^{-2(\eta\pi)^2\abs{i-j}} \hat{\sigma}_j^{-}(t).
\end{align} 
From these equations, one can calculate the mean-field equations \eqref{eq:MF} with the corresponding effective Rabi frequency \eqref{eq:effectiv_drive_EnsAve}. The output fields of the right-propagating mode just after the $N$-th emitter at position $z_N^+$ and of the left-propagating one just before the first emitter are given by
\begin{align}
    \hat{a}_\mathrm{out}^{\mathrm{EAM},R}(t) =& \hat{a}_\mathrm{in}^{R}(t) - i\sqrt{\frac{\Gamma_\mathrm{1D}}{2}}\sum_{i=1}^N\hat{\sigma}_i(t)\\
    \hat{a}_\mathrm{out}^{\mathrm{EAM},L}(t) =& \hat{a}_\mathrm{in}^{L}(t) - i\sqrt{\frac{\Gamma_\mathrm{1D}}{2}}\sum_{i=1}^N\hat{\sigma}_i(t)e^{-2(\eta\pi)^2\abs{i-1}}
\end{align}

\section{Continuum limit of unidirectional mean-field equations}\label{app:Continuos-limit}
In this section we show how to derive the continuous equation \eqref{eq:saturation_MF} from the mean-field equations. For simplicity we drop here the superindex UWM, since it is clear that we are dealing with the unidirectional waveguide model. The formal solutions for the emitter's means in dependence of the effective Rabi frequency are
\begin{align}\label{eq:MF_sol}
\expval*{\hat{\sigma}_i^-}=-2i\frac{\alpha_i\Gammatot}{\Gamma_\mathrm{tot}^2+8\alpha_i^2},\quad &\quad \expval{\hat{\sigma}_i^z}=-\frac{\Gamma_\mathrm{tot}^2}{\Gammatot^2+8\alpha_i^2}.
\end{align}
Taking now the difference equation $\Delta\alpha_i=\alpha_{i+1}-\alpha_i$ and inserting the solution for $\expval{\hat{\sigma}_i^-}$ imply a recursive relation for the effective Rabi frequency
\begin{align}\label{eq:differ-eq_alpha}
    \Delta\alpha_i&=\alpha_{i+1}-\alpha_i\nonumber\\
    &=-i\frac{\Gamma_\mathrm{1D}}{2}\expval*{\hat{\sigma}_i^-}\nonumber\\
    &=-2\beta\frac{\alpha_i}{1+8\frac{\alpha^2_i}{\Gammatot^2}},
\end{align}
where we used in the last step the relation $\beta=\Gamma_\mathrm{1D}/2\Gamma_\mathrm{tot}$. Now, we could take the continuous limit of Eq. \eqref{eq:differ-eq_alpha} arriving at a differential equation. However, we go one step further and derive a difference equation for $s_i=8\alpha_i^2/\Gamma_\mathrm{tot}^2$ before we take the continuous limit,
\begin{align}
    \Delta s_i&=s_{i+1}-s_i\nonumber\\
    &=\frac{8}{\Gamma_\mathrm{tot}^2}\left(\alpha_{i+1}^2-\alpha_i^2\right)\nonumber\\
    &=\frac{8}{\Gamma_\mathrm{tot}^2}\left(\alpha_{i+1}+\alpha_i\right)\left(\alpha_{i+1}-\alpha_i\right)\nonumber\\
    &\approx\frac{16}{\Gamma_\mathrm{tot}^2}\alpha_i\left(\alpha_{i+1}-\alpha_i\right)\nonumber\\
    &=-4\beta\frac{s_i}{1+s_i},
\end{align}
where we inserted Eq. \eqref{eq:differ-eq_alpha} in the last step. At this point we apply the substitution $i\rightarrow \ODi$ with $\ODi=4\beta i$ and take the continuous limit of the discrete index $\ODi$ getting the differential equation
\begin{align}
    \frac{ds(D)}{dD}=-\frac{s(D)}{1+s(D)}.
\end{align}
\end{document}